\documentclass[journal]{IEEEtran}
\usepackage{cite}
\usepackage{amsmath,amssymb,amsfonts}
\usepackage{graphicx}
\usepackage{array}
\usepackage{url}
\usepackage[hidelinks]{hyperref}
\newcommand{\norm}[1]{\left\lVert#1\right\rVert}
\newcommand{\E}{\mathbb{E}}
\newcommand{\Real}{\operatorname{Re}}
\newcommand{\Imag}{\operatorname{Im}}
\begin{document}
\title{Optical Coherent Combining in Multi-Aperture Free-Space Receivers through Atmospheric Turbulence via Optically Computed Alignment Gradients}
\author{Chen~Cheng, Tong~Luo, Siyu~Gong, Qun~Zhang, Linsheng~Fan, Qi~Wu, Yanfu~Yang, and Jiayin~Xue%
\thanks{C.~Cheng, T.~Luo, S.~Gong, Q.~Zhang, L.~Fan, Q.~Wu, and J.~Xue are with Pengcheng Laboratory, Shenzhen 518000, China.}%
\thanks{Y.~Yang is with the Department of Electronic and Information Engineering, Harbin Institute of Technology, Shenzhen 518055, China (e-mail: yangyanfu@hit.edu.cn).}%
\thanks{Corresponding author: Jiayin~Xue (e-mail: xuejy@pcl.ac.cn).}%
\thanks{Working preprint prepared for arXiv. Funding and disclosure clearance must be confirmed before public submission.}}
\markboth{Working Preprint --- Not Yet Submitted}{Optical Coherent Combining via Optically Computed Alignment Gradients}
\maketitle
\begin{abstract}
Multi-aperture reception turns spatial diversity into a coherent-combining gain only if the field replicas collected by the apertures are aligned in phase and polarization before they are summed, and in a satellite-to-ground downlink the alignment has to track the atmosphere. This paper shows that the exact gradients of the combined output power with respect to the phase and the polarization angle of every branch can be measured optically, before photodetection. A passive tap of the combined field, split one-to-$N$, serves as a common self-reference. Interfered with a tap of branch $i$ on a 3~dB coupler and detected by a balanced detector, it yields the phase gradient $-\partial P_{\rm out}/\partial\varphi_i=2\Imag\{\mathbf{s}^{H}\mathbf{v}_i\}$; the same coupler, preceded on the branch arm by a $90^{\circ}$ Faraday rotator and a static $\pi/2$ offset, yields the polarization gradient $\partial P_{\rm out}/\partial\theta_i=2\Real\{\mathbf{s}^{H}\mathbf{J}\mathbf{v}_i\}$. Each branch therefore carries a phase loop and a polarization loop that update simultaneously, with fixed gain and without division, perturbation, reference laser, or per-branch coherent receiver, and the loops can be analog integrators or a low-rate digital processor. The paper derives the gradients and their readout from elementary optical parts, shows geometrically how the two updates turn each branch toward the combined field and why the phase loop has to lock first, and gives the loop bandwidth, its stability condition and a one-time design rule. A phase-only simulation compares the digital loop with centralized and distributed stochastic parallel gradient descent and with dither locking under identical disturbances, a common 100-MS/s clock, the same front-end amplifier for every beam, and tuning of every scheme for every condition, at a per-beam OSNR of 6~dB, where ideal combining of four beams just reaches the FEC threshold of a 28-GBd DP-QPSK signal. For a multi-tone phase disturbance with rms rates from 10 to $10^{5}$~rad/s, where the clock does not limit the proposed loop, the loop leaves the lowest residual loss of the four---at most $5.6\times10^{-4}$~dB for four beams, 13 to 160 times less than the other schemes---reaches 90\% efficiency within 0.6~$\mu$s for up to 64 beams, 2.8 to 59 times faster than the others wherever they hold the lock (centralized stochastic parallel gradient descent loses it at 64 beams and $10^{5}$~rad/s), and needs only the single step size that the design rule predicts; it needs $2N$ photodiodes but no perturbation or dither frequency. Between per-beam OSNRs of $-10$ and 10~dB at a constant amplifier output, the loop loses 8 to 100 times less than the other schemes, which fail about 10~dB earlier when the OSNR falls further, and raising the output beyond 0~dBm per beam changes nothing. For the DP-QPSK signal, the readout taps leave the OSNR at the receiver unchanged and every retuned scheme stays within 0.06~dB of ideal combining at the FEC threshold, as a symbol-level simulation with carrier recovery confirms, whereas brief misalignments of dither locking cost it 0.14~dB at an OSNR of 10~dB and $10^{5}$~rad/s and, with the settings held fixed, 1.0~dB at the threshold. Up to $10^{7}$~rad/s, 1~dB above the threshold, where the clock limits every scheme, the loop keeps the bit-error ratio of ideal combining for four, sixteen, and sixty-four beams, whereas every baseline crosses the threshold between $10^{5}$ and $10^{7}$~rad/s, the earlier the more beams, and, in a simulation of the complete chain, makes the carrier recovery slip. In fades with the settings held fixed, the other schemes lose 0.2 to 0.9~dB at a per-beam OSNR of $-14$~dB, whereas the loop, whose bandwidth falls with the received power, loses $4\times10^{-3}$~dB. The assumptions and limitations are stated explicitly; experimental validation remains open.
\end{abstract}

\section{Introduction}
\IEEEPARstart{M}{ulti-aperture} reception is a practical route to spatial diversity in free-space optical (FSO) links. In a satellite-to-ground downlink each aperture collects a partially independent copy of the same field, perturbed in amplitude and phase by atmospheric turbulence and rotated in polarization by the tracking optics and by the aperture-specific fiber coupling; summing the copies coherently restores the received power and mitigates deep fades~\cite{Khalighi2014,Guiomar2022}. The copies add constructively only if they are aligned in phase and polarization, so a multi-aperture receiver is defined largely by where the alignment and the summation take place and by what observation drives the alignment.

In the dominant architecture every aperture has its own coherent receiver, digitizer, and digital alignment stage~\cite{Geisler2016,Ip2008,Liu2023}, so the number of high-speed receivers, analog-to-digital converters, and complex-arithmetic channels grows with the aperture count $N$. Combining optically, before photodetection, leaves a single coherent receiver for the combined field. Reported optical combiners, however, drive their actuators from scalar power measurements or dedicated probes~\cite{Yang2017,Billault2021,Larsson2022}: a per-parameter update direction is then obtained by perturbing the parameters, by modulating them at distinct frequencies, or by sweeping, and not from a single measurement.

The scheme of this paper starts from a digital-domain lineage. Blind gradient-ascent phase alignment (BGAPA) maximizes the combined output power with a closed-form gradient~\cite{Chen2026a}, and the heterogeneous-gradient formulation extends the same objective to the polarization angle of each aperture~\cite{Chen2026}. In both, the update direction of each parameter is an inner-product component between the combined field and the branch field---an exact gradient of the total power, but computed digitally after per-aperture detection. These inner-product components are exactly what a balanced interferometer measures, so the power gradients can be read out optically, before photodetection and without digitizing the control path. The contributions are the following.
\begin{enumerate}
\item An architecture in which a passive tap of the combined field, split one-to-$N$, provides one self-reference per branch, so that $N$ readout units produce all phase and polarization gradients at once while the data path keeps a single coherent receiver (Section~\ref{sec:model}).
\item The exact gradients $\partial P_{\rm out}/\partial\varphi_i=-2\Imag\{\mathbf{s}^{H}\mathbf{v}_i\}$ and $\partial P_{\rm out}/\partial\theta_i=2\Real\{\mathbf{s}^{H}\mathbf{J}\mathbf{v}_i\}$; their readout with a 3~dB coupler, a static phase offset, a $90^{\circ}$ Faraday rotator, and balanced detectors; and the geometry of the two updates, which shows how each loop turns its branch toward the combined field, how the two loops are coupled, and why the phase loop has to lock first (Section~\ref{sec:readout}).
\item Fixed-gain, division-free loops, realized by analog integrators or by a low-rate processor, with their bandwidth, stability condition, and a one-time design rule, together with the assumptions and limitations of the scheme (Sections~\ref{sec:readout} and~\ref{sec:limits}).
\item A simulated comparison, restricted to the phase, with centralized and distributed stochastic parallel gradient descent and with dither locking, under identical disturbances, the same sampling rate of 100~MS/s, the same front-end amplification, and tuning of every scheme for every condition, at the OSNR at which ideal combining reaches the FEC threshold and over phase rates up to $10^{5}$~rad/s, array sizes, the OSNR at constant and at falling power, and the amplifier output power, together with the bit-error ratio of a DP-QPSK signal up to $10^{7}$~rad/s for four, sixteen, and sixty-four beams, also from a simulation of the complete chain; it confirms the design rule for the step size of the digital loop, shows that the readout taps leave the OSNR at the receiver unchanged, and counts the control-path hardware of every scheme versus $N$ (Section~\ref{sec:compare}).
\end{enumerate}

\section{Combining Architecture and Signal Model}
\label{sec:model}
\subsection{Receiver Architecture}
\begin{figure*}[t]
\centering\includegraphics[width=\textwidth]{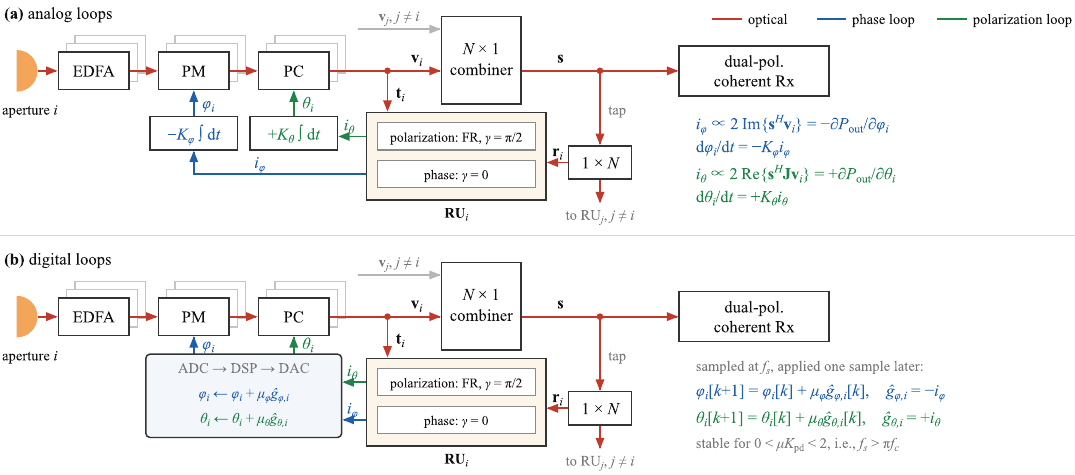}
\caption{Branch $i$ of the $N$-branch optical-domain combiner with its gradient readout and its two fixed-gain loops, in the two realizations of the loops (red: optical; blue: phase loop; green: polarization loop). The other $N-1$ branches are identical (stacked boxes) and are combined with $\mathbf{v}_i$. A tap $\mathbf{t}_i$ of the compensated branch field $\mathbf{v}_i$ interferes in the readout unit RU$_i$ (Fig.~\ref{fig:readout}) with a self-reference $\mathbf{r}_i$, split $1\times N$ from a tap of the combined field $\mathbf{s}$. The phase readout returns $i_{\varphi}\propto-\partial P_{\rm out}/\partial\varphi_i$, which drives the phase modulator (PM); the polarization readout returns $i_{\theta}\propto+\partial P_{\rm out}/\partial\theta_i$, which drives the polarization controller (PC). (a)~Analog loops: integrators accumulate the photocurrents continuously, $d\varphi_i/dt=-K_{\varphi}i_{\varphi}$ and $d\theta_i/dt=+K_{\theta}i_{\theta}$ [Fig.~\ref{fig:loopimpl}(a)]. (b)~Digital loops: an ADC samples both readouts at $f_s$, a processor takes the steps $\varphi_i\leftarrow\varphi_i+\mu_{\varphi}\hat{g}_{\varphi,i}$ and $\theta_i\leftarrow\theta_i+\mu_{\theta}\hat{g}_{\theta,i}$ along the measured ascent directions $\hat{g}_{\varphi,i}\propto-i_{\varphi}$ and $\hat{g}_{\theta,i}\propto+i_{\theta}$, and a DAC applies them one sample later; the loops are stable for $0<\mu K_{\rm pd}<2$ [Eqs.~\eqref{eq:dspphi}--\eqref{eq:dsptheta}, Fig.~\ref{fig:loopimpl}(b)]. The EDFA runs at fixed gain and the branch has no amplitude actuator; all loops run in parallel and never interrupt the data path to the single coherent receiver. FR: Faraday rotator.}
\label{fig:architecture}
\end{figure*}

The receiver combines $N\geq2$ copies of the same dual-polarization signal, collected by spatially separated apertures, photonic-lantern mode ports, or multicore-fiber cores. Fig.~\ref{fig:architecture} shows branch $i$; the other $N-1$ branches are identical. The branch is amplified by an EDFA at \emph{fixed gain}---without automatic gain control, so that the branches keep their scintillation-induced amplitude differences and every branch is combined at its own field amplitude (Section~\ref{sec:limits})---and then passes through a phase modulator (PM) and a polarization controller (PC); there is no amplitude actuator. The compensated branch fields $\mathbf{v}_i$ are summed passively in an $N\times1$ combiner, whose output $\mathbf{s}$ goes to a single dual-polarization coherent receiver. A small tap of $\mathbf{s}$, split one-to-$N$, provides a self-reference $\mathbf{r}_i\propto\mathbf{s}$ to the readout unit RU$_i$ of every branch, where it interferes with a tap $\mathbf{t}_i\propto\mathbf{v}_i$ of the compensated branch. RU$_i$ returns two differential photocurrents: $i_{\varphi}$, proportional to the phase gradient, closes the phase loop on the PM, and $i_{\theta}$, proportional to the polarization gradient, closes the polarization loop on the PC. Each loop is a single accumulator with a fixed gain, realized either by an analog integrator [Fig.~\ref{fig:architecture}(a)] or by a low-rate digital processor that steps each actuator along the sampled readout, $\varphi_i\leftarrow\varphi_i+\mu_{\varphi}\hat{g}_{\varphi,i}$ [Fig.~\ref{fig:architecture}(b)]; the two realizations are equivalent at small step size (Section~\ref{subsec:loopimpl}). All $2N$ loops run in parallel, and none of them interrupts the data path.

Three features distinguish this readout from a scalar power monitor. The observable is the difference of two balanced ports, so the self-powers of the fields cancel. It is branch-resolved, so all $N$ gradients are available in the same averaging window, without frequency, code, or time multiplexing. And the reference is a tap of the combined field itself, so no separate laser or optical frequency lock is needed, and the reference carries the same carrier and modulation as every branch.

\subsection{Combining Network}
The branches are summed by a binary tree of $\log_2N$ stages of $2\times2$ 3~dB couplers. Each stage adds an excess loss $L_{\rm ex}$ to the ideal splitting, so the cascade attenuates the coherent sum by $\log_2(N)L_{\rm ex}$ in decibels, which grows only logarithmically with $N$; this loss, like the efficiency penalty of a residual phase error, is common to all optical combiners and is not treated further. A binary tree fixes only the relative phase of each pair, so an $N$-input combiner has $N-1$ phase degrees of freedom; the branch-resolved loops actuate one phase per branch, which spans these $N-1$ dimensions plus an unobservable common mode. Because every stage is passive, the power available at the monitoring tap is a fixed fraction of the combined power and grows with the array gain rather than falling with the number of branches.

\subsection{Signal Model}
All fields are dual-polarization Jones vectors in a common linear basis. Branch $i$ before compensation is $\mathbf{v}_i^{(0)}$; after the PM and the PC it is
\begin{equation}
\mathbf{v}_i[k]=\mathbf{R}(\theta_i[k])\,e^{j\varphi_i[k]}\,\mathbf{v}_i^{(0)}[k],
\label{eq:model}
\end{equation}
where $k$ indexes the control window and
\begin{equation}
\mathbf{R}(\theta)=\begin{bmatrix}\cos\theta&\sin\theta\\-\sin\theta&\cos\theta\end{bmatrix}
\label{eq:R}
\end{equation}
rotates the direction of polarization by $-\theta$. The combiner sums the compensated branches,
\begin{equation}
\mathbf{s}[k]=\sum_{i=1}^{N}\mathbf{v}_i[k];
\label{eq:s}
\end{equation}
its common field scale ($1/\sqrt{N}$ for an ideal combiner) is immaterial to the gradients, and there is no per-branch amplitude weight. Equation~\eqref{eq:model} is a deliberately restricted channel family: the differential impairments between branches are a scalar phase and a real polarization rotation, which capture the dominant turbulence- and fiber-induced misalignments in the intended use. General complex Jones matrices, polarization-dependent loss, and channel memory are outside the model (Section~\ref{sec:limits}). Timing between branches is assumed pre-aligned, so that within a control window the data waveforms of $\mathbf{s}$ and $\mathbf{v}_i$ remain correlated. Within a window the actuator settings are held, and every readout unit sees the combined field as it is, including the contribution of its own branch.

\section{Optical Readout of the Alignment Gradients}
\label{sec:readout}
\subsection{Output-Power Objective and Its Exact Gradients}
The objective is the total power delivered by the combiner,
\begin{equation}
P_{\rm out}[k]=\norm{\mathbf{s}[k]}^{2}=|s_x[k]|^{2}+|s_y[k]|^{2},
\label{eq:loss}
\end{equation}
to be maximized over the phase $\varphi_i$ and the polarization angle $\theta_i$ of every branch. Both polarization components of $\mathbf{s}=[s_x,s_y]^{T}$ enter, so the objective is the total power and not the power in one polarization.

Because $\mathbf{s}=\sum_j\mathbf{v}_j$ and the parameters of branch $i$ enter only through $\mathbf{v}_i$, the derivative with respect to $\xi\in\{\varphi_i,\theta_i\}$ is $\partial_\xi\norm{\mathbf{s}}^{2}=2\Real\{\mathbf{s}^{H}\partial_\xi\mathbf{v}_i\}$, evaluated at the current combined field. The model~\eqref{eq:model} gives $\partial_{\varphi_i}\mathbf{v}_i=j\mathbf{v}_i$, because the PM multiplies both components by $e^{j\varphi_i}$, and $\partial_{\theta_i}\mathbf{v}_i=\mathbf{J}\mathbf{v}_i$, because the rotation mixes the two components without scaling them, with
\begin{equation}
\mathbf{J}=\begin{bmatrix}0&1\\-1&0\end{bmatrix},\qquad\mathbf{J}\mathbf{v}_i=[v_{i,y},\,-v_{i,x}]^{T}.
\label{eq:J}
\end{equation}
Using $\Real\{jz\}=-\Imag\{z\}$,
\begin{align}
\frac{\partial P_{\rm out}}{\partial\varphi_i}
&=-2\,\Imag\{\mathbf{s}^{H}\mathbf{v}_i\}=-2\,\Imag\bigl\{s_x^{*}v_{i,x}+s_y^{*}v_{i,y}\bigr\},\label{eq:gphi}\\
\frac{\partial P_{\rm out}}{\partial\theta_i}
&=\;\;2\,\Real\{\mathbf{s}^{H}\mathbf{J}\mathbf{v}_i\}=\;\;2\,\Real\bigl\{s_x^{*}v_{i,y}-s_y^{*}v_{i,x}\bigr\}.\label{eq:gtheta}
\end{align}
These are the ascent directions of the digital heterogeneous-gradient formulation~\cite{Chen2026}, where they carry a branch weight $p_n$ (its Eqs.~(10) and~(11)); without an amplitude weight that factor is absorbed into the loop gain, and the polarization sums in~\eqref{eq:gphi} and~\eqref{eq:gtheta} are formed by the detectors instead of in digital arithmetic.

Three properties make the gradients usable as control laws. They are \emph{exact}: each is the true gradient of the delivered power at the current operating point, so no frozen reference, perturbation, search, or training sequence is needed. They are \emph{branch-resolved}: each involves only the pair $(\mathbf{s},\mathbf{v}_i)$, so all branches can be updated simultaneously, as a Jacobi iteration. And only their \emph{direction} matters: a common positive factor (splitting ratios, reference attenuation, responsivity, transimpedance) is absorbed into the loop gain, so the loops need neither normalization nor division. The objective itself is the power---the same scalar that perturbation-based combiners maximize; the scheme differs from them in how the update direction is obtained, as a correlation measured in one shot rather than as a finite difference of power readings.

\subsection{Geometry of the Two Updates}
\label{subsec:geometry}
\begin{figure}[t]
\centering\includegraphics[width=\columnwidth]{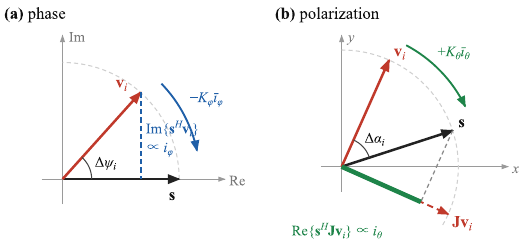}
\caption{Geometry of the two gradients and of one update step, for linearly polarized fields. (a) Phase: in the complex plane of the readout, $\Imag\{\mathbf{s}^{H}\mathbf{v}_i\}$ is the component of $\mathbf{v}_i$ in quadrature to $\mathbf{s}$, and the step $-K_{\varphi}\bar{i}_{\varphi}$ turns $\mathbf{v}_i$ toward $\mathbf{s}$. (b) Polarization: $\mathbf{J}\mathbf{v}_i$ is $\mathbf{v}_i$ rotated by $-90^{\circ}$, $\Real\{\mathbf{s}^{H}\mathbf{J}\mathbf{v}_i\}$ is the projection of $\mathbf{s}$ on it, and the step $+K_{\theta}\bar{i}_{\theta}$ rotates the polarization of $\mathbf{v}_i$ toward that of $\mathbf{s}$. Both readouts vanish, and both loops come to rest, when $\mathbf{v}_i$ is aligned with $\mathbf{s}$; Eq.~\eqref{eq:geom} gives their magnitudes.}
\label{fig:geometry}
\end{figure}

The gradients take a transparent form for linearly polarized fields, the case in which the real rotation~\eqref{eq:R} acts. Write $\mathbf{s}=\norm{\mathbf{s}}\,\hat{\mathbf{e}}(\alpha_s)\,e^{j\psi_s}$ and $\mathbf{v}_i=\norm{\mathbf{v}_i}\,\hat{\mathbf{e}}(\alpha_i)\,e^{j\psi_i}$ with $\hat{\mathbf{e}}(\alpha)=[\cos\alpha,\sin\alpha]^{T}$, and let $\Delta\psi_i=\psi_i-\psi_s$ and $\Delta\alpha_i=\alpha_i-\alpha_s$ be the phase and polarization misalignments of branch $i$ with respect to the combined field. Then $\mathbf{s}^{H}\mathbf{v}_i=\norm{\mathbf{s}}\norm{\mathbf{v}_i}\cos\Delta\alpha_i\,e^{j\Delta\psi_i}$ and $\mathbf{s}^{H}\mathbf{J}\mathbf{v}_i=\norm{\mathbf{s}}\norm{\mathbf{v}_i}\sin\Delta\alpha_i\,e^{j\Delta\psi_i}$, so that
\begin{equation}
\begin{split}
\frac{\partial P_{\rm out}}{\partial\varphi_i}&=-2\norm{\mathbf{s}}\norm{\mathbf{v}_i}\cos\Delta\alpha_i\,\sin\Delta\psi_i,\\
\frac{\partial P_{\rm out}}{\partial\theta_i}&=+2\norm{\mathbf{s}}\norm{\mathbf{v}_i}\sin\Delta\alpha_i\,\cos\Delta\psi_i.
\end{split}
\label{eq:geom}
\end{equation}
Each gradient is the sine of its own misalignment, weighted by the cosine of the other misalignment and by the product of the branch and combined field amplitudes. In the complex plane of the phase readout [Fig.~\ref{fig:geometry}(a)], $\Imag\{\mathbf{s}^{H}\mathbf{v}_i\}$ is the component of $\mathbf{v}_i$ in quadrature to $\mathbf{s}$, and the phase step $-K_{\varphi}\,\bar{i}_{\varphi}$ of Section~\ref{subsec:updates} turns $\mathbf{v}_i$ toward $\mathbf{s}$ until that component vanishes. In the plane of polarization [Fig.~\ref{fig:geometry}(b)], $\mathbf{J}\mathbf{v}_i$ is $\mathbf{v}_i$ rotated by $-90^{\circ}$ and $\Real\{\mathbf{s}^{H}\mathbf{J}\mathbf{v}_i\}$ is the projection of $\mathbf{s}$ on it; the step $+K_{\theta}\,\bar{i}_{\theta}$, which by~\eqref{eq:R} rotates the polarization of $\mathbf{v}_i$ by $-K_{\theta}\,\bar{i}_{\theta}$, turns it toward the polarization of $\mathbf{s}$.

Three consequences for the joint update follow from~\eqref{eq:geom}. Near alignment the cosines are close to one, so the gradients reduce to $-2\norm{\mathbf{s}}\norm{\mathbf{v}_i}\Delta\psi_i$ and $2\norm{\mathbf{s}}\norm{\mathbf{v}_i}\Delta\alpha_i$: each branch runs two independent first-order loops of the same form, whose rate is proportional to the field product (Section~\ref{subsec:loopimpl}). Far from alignment the loops are coupled through the cosines; in particular, the polarization gradient points the wrong way while $|\Delta\psi_i|>\pi/2$, which is why the phase loop is closed first (Section~\ref{subsec:acq}). And a branch whose polarization starts more than $90^{\circ}$ away from that of the combined field is not lost: the phase loop then settles at $\Delta\psi_i=\pi$ and the polarization loop at $\Delta\alpha_i=\pi$, which describe the same field as $\Delta\psi_i=\Delta\alpha_i=0$, because $\hat{\mathbf{e}}(\alpha+\pi)\,e^{j(\psi+\pi)}=\hat{\mathbf{e}}(\alpha)\,e^{j\psi}$.

\subsection{Phase-Gradient Readout}
\label{subsec:phasereadout}
A directional coupler is polarization insensitive, so the same scalar port matrix acts on both Jones components and can be written for the Jones vectors. With symmetric reference planes, and with common scale factors dropped, a 3~dB coupler fed with the reference $\mathbf{s}$ and the branch tap $\mathbf{v}_i$ delivers
\begin{equation}
\begin{bmatrix}\mathbf{p}\\ \mathbf{q}\end{bmatrix}
=\frac{1}{\sqrt{2}}\begin{bmatrix}1&-j\\-j&1\end{bmatrix}
\begin{bmatrix}\mathbf{s}\\ \mathbf{v}_i\end{bmatrix},
\label{eq:ports}
\end{equation}
and square-law detection of the two ports gives
\begin{equation}
P(\mathbf{p}),\,P(\mathbf{q})=\frac{\norm{\mathbf{s}}^{2}+\norm{\mathbf{v}_i}^{2}}{2}\pm\Imag\{\mathbf{s}^{H}\mathbf{v}_i\}.
\label{eq:portpowers}
\end{equation}
The squared modulus itself produces the conjugate, and the balanced difference cancels the self-powers:
\begin{equation}
i\triangleq P(\mathbf{p})-P(\mathbf{q})=2\,\Imag\{\mathbf{s}^{H}\mathbf{v}_i\}.
\label{eq:balance}
\end{equation}
A bare 3~dB coupler is therefore a quadrature extractor, because its through and cross paths differ by $\pi/2$. Since one scalar matrix acts on both polarizations, the two components are detected and summed by the same photodiodes, $\Imag\{\mathbf{s}^{H}\mathbf{v}_i\}=\Imag\{s_x^{*}v_{i,x}\}+\Imag\{s_y^{*}v_{i,y}\}$, and no polarization beam splitter is needed; detecting one polarization alone would not be equivalent.

A static, polarization-independent phase offset $\gamma$ on the branch arm interpolates between the two quadratures,
\begin{equation}
\begin{split}
i(\gamma)&=2\,\Imag\{\mathbf{s}^{H}e^{j\gamma}\mathbf{v}_i\}\\
&=\cos\gamma\,2\,\Imag\{\mathbf{s}^{H}\mathbf{v}_i\}+\sin\gamma\,2\,\Real\{\mathbf{s}^{H}\mathbf{v}_i\},
\end{split}
\label{eq:gammaselect}
\end{equation}
so $\gamma=0$ keeps the quadrature and $\gamma=\pi/2$ selects the in-phase component. The offset is the elementary-parts equivalent of the internal $\pi/2$ that separates the I and Q ports of a $90^{\circ}$ hybrid, and it also trims the residual differential path length of its readout unit (Section~\ref{sec:limits}). With $\gamma=0$, by~\eqref{eq:gphi},
\begin{equation}
i_{\varphi}=2\,\Imag\{\mathbf{s}^{H}\mathbf{v}_i\}=-\frac{\partial P_{\rm out}}{\partial\varphi_i}:
\label{eq:iphi}
\end{equation}
the coupler and the balanced detector alone measure the phase gradient (Fig.~\ref{fig:readout}, top). Reading this quadrature rather than a dark port keeps the readout linear at lock: by~\eqref{eq:geom} it is proportional to $\sin\Delta\psi_i$ rather than to $1-\cos\Delta\psi_i$, so it changes sign with the error, and a branch at the anti-phase equilibrium is driven away from it. For a coupling angle $\kappa\neq\pi/4$, \eqref{eq:balance} becomes $\cos2\kappa\,(\norm{\mathbf{s}}^{2}-\norm{\mathbf{v}_i}^{2})+2\sin2\kappa\,\Imag\{\mathbf{s}^{H}\mathbf{v}_i\}$; the leaked self-power term offsets the lock point in proportion to the splitting error and is calibrated out with the detector offsets (Section~\ref{sec:limits}).

\subsection{Polarization-Gradient Readout}
\label{subsec:polreadout}
The polarization gradient~\eqref{eq:gtheta} carries $\mathbf{J}$ instead of the factor $j$: it is not a quadrature of $\mathbf{s}^{H}\mathbf{v}_i$ but the in-phase component of $\mathbf{s}^{H}\mathbf{J}\mathbf{v}_i$. Two elementary operations produce it from the same taps. The branch tap is rotated by $90^{\circ}$ in an in-line Faraday rotator, whose Jones matrix is a genuine rotation,
\begin{equation}
\mathbf{F}(\beta)=\begin{bmatrix}\cos\beta&-\sin\beta\\ \sin\beta&\cos\beta\end{bmatrix},
\qquad
\mathbf{F}(90^{\circ})=-\mathbf{J},
\label{eq:faraday}
\end{equation}
and the in-phase component is selected by the offset $\gamma=\pi/2$ of~\eqref{eq:gammaselect}. The balanced difference is then
\begin{equation}
i_{\theta}=2\,\Real\{\mathbf{s}^{H}\mathbf{J}\mathbf{v}_i\}=+\frac{\partial P_{\rm out}}{\partial\theta_i},
\label{eq:readoutgrad}
\end{equation}
up to a sign that the detector polarity absorbs at calibration (Fig.~\ref{fig:readout}, bottom).

The rotation has to be a real polarization conversion: orthogonally polarized fields do not beat in a polarization-insensitive coupler, so merely cross-connecting orthogonal ports would give no signal. A single half-wave plate cannot replace the rotator, because a retarder has a symmetric Jones matrix whereas $\mathbf{J}$ is antisymmetric; two static half-wave plates at $0^{\circ}$ and $45^{\circ}$ give $\pm\mathbf{J}$ and are an alternative, as is a polarization splitter whose $x$ and $y$ outputs are swapped, one of them with a $\pi$ shift, and recombined. The non-reciprocity of the rotator is admissible because it sits in the one-way monitoring path. A rotator error $\epsilon=\beta-90^{\circ}$ gives $\mathbf{F}=-\cos\epsilon\,\mathbf{J}-\sin\epsilon\,\mathbf{I}$ and leaks $\sin\epsilon\cdot2\Real\{\mathbf{s}^{H}\mathbf{v}_i\}$ into $i_{\theta}$, $1.7\%$ per degree of error; like the coupler error, it is static and is calibrated out. The $90^{\circ}$ Jones rotation of the polarization vector and the $90^{\circ}$ optical hybrid of I/Q detection are distinct operations: the first acts on the polarization, the second on the carrier phase.

\subsection{Readout Unit and Update Laws}
\label{subsec:updates}
\begin{figure}[t]
\centering\includegraphics[width=\columnwidth]{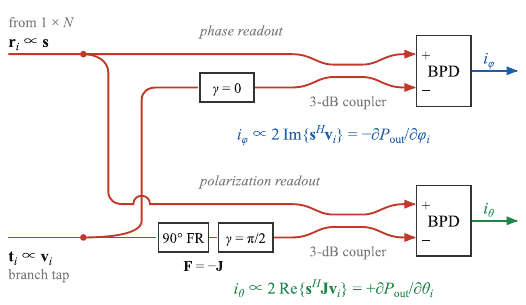}
\caption{Readout unit RU$_i$ of Fig.~\ref{fig:architecture}. The self-reference $\mathbf{r}_i\propto\mathbf{s}$ and the branch tap $\mathbf{t}_i\propto\mathbf{v}_i$ are each split in two. Top: with the static offset at $\gamma=0$, the 3~dB coupler and the balanced photodetector (BPD) deliver the quadrature $i_{\varphi}\propto2\Imag\{\mathbf{s}^{H}\mathbf{v}_i\}=-\partial P_{\rm out}/\partial\varphi_i$. Bottom: the $90^{\circ}$ Faraday rotator (FR) applies $\mathbf{F}=-\mathbf{J}$ to the branch arm and the offset $\gamma=\pi/2$ selects the in-phase port, so the second BPD delivers $i_{\theta}\propto2\Real\{\mathbf{s}^{H}\mathbf{J}\mathbf{v}_i\}=+\partial P_{\rm out}/\partial\theta_i$. Every waveguide carries both polarizations and each BPD sums them, so no polarization beam splitter is needed; the crossing fibers do not connect, and detector polarities are fixed at calibration.}
\label{fig:readout}
\end{figure}

Fig.~\ref{fig:readout} assembles both readouts into RU$_i$. The reference and the branch tap are each split in two: one pair interferes on a coupler with $\gamma=0$ and gives $i_{\varphi}$; in the other pair the branch arm passes the Faraday rotator and the offset $\gamma=\pi/2$, and the second coupler gives $i_{\theta}$. Collecting splitting ratios, responsivities, transimpedances, and actuator slopes into two calibrated loop gains $K_{\varphi}$ and $K_{\theta}$, with polarities fixed at calibration, the updates are
\begin{align}
\varphi_i[k+1]&=\varphi_i[k]-K_{\varphi}\,\bar{i}_{\varphi}[k],
\label{eq:philoop}\\
\theta_i[k+1]&=\theta_i[k]+K_{\theta}\,\bar{i}_{\theta}[k],
\label{eq:thetaloop}
\end{align}
where the overbar denotes the low-pass-filtered readout. By~\eqref{eq:iphi} and~\eqref{eq:readoutgrad} both are gradient-ascent steps on $P_{\rm out}$, and all branches take them at once. They need no normalization, division, or regularization floor: the gains are calibrated once, at the maximum working optical power, and at run time the loops only amplify and accumulate differential currents. The digital formulations~\cite{Chen2026,Chen2026a} perform the same constant-step ascent from digitized samples; here the dependence of the effective step on the branch and reference powers, visible in~\eqref{eq:geom}, is handled by gain headroom alone, since there is no amplitude actuator (Section~\ref{sec:limits}).

\subsection{Loop Realization, Bandwidth, and Design Rule}
\label{subsec:loopimpl}
\begin{figure}[t]
\centering\includegraphics[width=\columnwidth]{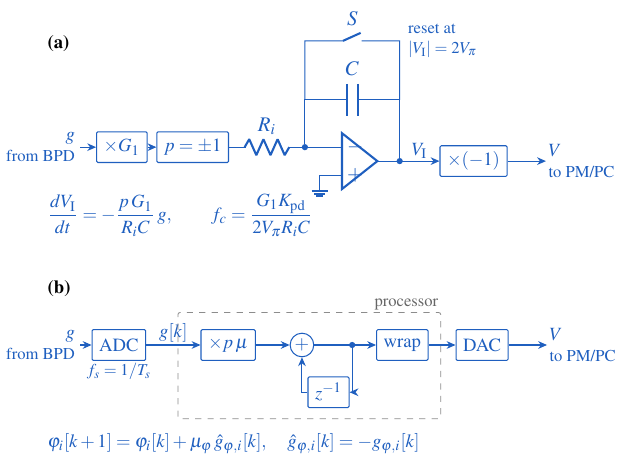}
\caption{Two realizations of one fixed-gain loop (phase or polarization) of Fig.~\ref{fig:architecture}. (a) Analog: preamplifier $G_1$, calibrated polarity $p$, operational-amplifier integrator $R_iC$ with reset switch $S$, and inverting driver, giving~\eqref{eq:analogint} and the bandwidth~\eqref{eq:fc}. (b) Digital: ADC at $f_s$, a processor that scales by $p\mu$, accumulates ($z^{-1}$), and wraps, and a DAC holding the actuator voltage, giving~\eqref{eq:dspphi}--\eqref{eq:dsptheta}. $g$: BPD output voltage of the loop.}
\label{fig:loopimpl}
\end{figure}

Each of the $2N$ loops~\eqref{eq:philoop}--\eqref{eq:thetaloop} needs only an accumulator between its balanced detector and its actuator (Fig.~\ref{fig:loopimpl}). The gradient itself is formed optically, so the electronics handle one real, low-rate signal per loop, at the loop bandwidth rather than at the symbol rate. Let $g$ be the detector output voltage of a loop ($i_{\varphi}$ or $i_{\theta}$ times the transimpedance), $K_{\rm pd}$ its slope with respect to the alignment error at lock (V/rad), $V_\pi$ the actuator half-wave voltage, and $p=\pm1$ the polarity fixed at calibration. Since $g_{\varphi}\propto-\partial P_{\rm out}/\partial\varphi_i$ and $g_{\theta}\propto+\partial P_{\rm out}/\partial\theta_i$, ascent requires $p=-1$ for the phase loop and $p=+1$ for the polarization loop, up to the detector polarity.

\emph{Analog integrator.} The output $g$ is amplified by $G_1$, given the polarity $p$, and integrated by an operational-amplifier integrator of time constant $R_iC$; an inverting driver delivers $V=-V_{\rm I}$ to the actuator [Fig.~\ref{fig:loopimpl}(a)]:
\begin{equation}
\frac{dV_{\rm I}}{dt}=-\frac{p\,G_1}{R_iC}\,g.
\label{eq:analogint}
\end{equation}
This is the continuous-time form of~\eqref{eq:philoop}--\eqref{eq:thetaloop}: the charge on $C$ is the accumulator and $G_1/(R_iC)$ the step size. With the actuator slope $\pi/V_\pi$, the loop linearized at lock is a pure integrator of bandwidth
\begin{equation}
f_c=\frac{G_1K_{\rm pd}}{2V_\pi R_iC}.
\label{eq:fc}
\end{equation}
An input-referred offset $V_{\rm os}$ moves the lock point by $V_{\rm os}/K_{\rm pd}$ and is trimmed at calibration. The integrator output is bounded by the supply rails, whereas the drift it tracks is not: a window comparator closes the switch $S$ when $|V_{\rm I}|$ reaches $2V_\pi$, discharging $C$ and stepping the phase by exactly $2\pi$, an equivalent lock point, so the reset does not disturb the combined field. The polarization loop is reset in the same way, by a $2\pi$ rotation of $\theta_i$ or by $\pi$ together with a $\pi$ step of $\varphi_i$, since $\mathbf{R}(\theta+\pi)=-\mathbf{R}(\theta)$.

\emph{Digital processor.} An analog-to-digital converter samples $g$ every $T_s=1/f_s$, a processor accumulates, and a digital-to-analog converter holds the actuator voltage until the next sample [Fig.~\ref{fig:loopimpl}(b)]:
\begin{align}
\varphi_i[k+1]&=\varphi_i[k]+\mu_\varphi\,\hat{g}_{\varphi,i}[k],\qquad \hat{g}_{\varphi,i}[k]=-g_{\varphi,i}[k],
\label{eq:dspphi}\\
\theta_i[k+1]&=\theta_i[k]+\mu_\theta\,\hat{g}_{\theta,i}[k],\qquad \hat{g}_{\theta,i}[k]=+g_{\theta,i}[k],
\label{eq:dsptheta}
\end{align}
which are~\eqref{eq:philoop}--\eqref{eq:thetaloop} with $K_\varphi$ and $K_\theta$ realized as the step sizes $\mu_\varphi$ and $\mu_\theta$. Linearized at lock, and with the new setting applied one sample after the measurement, the alignment error obeys $e[k+1]=(1-\mu K_{\rm pd})\,e[k]$ in the absence of disturbance: each sample corrects the fraction $\mu K_{\rm pd}=2\pi f_cT_s$ of the error, so the loop is stable for $0<\mu K_{\rm pd}<2$, i.e., for $f_s>\pi f_c$, and it matches the analog integrator for $\mu K_{\rm pd}\ll1$, with $\mu=\pi G_1T_s/(V_\pi R_iC)$. The phase accumulator is wrapped into $[-\pi,\pi)$, the digital counterpart of the $2V_\pi$ reset. Unlike digital per-aperture combining, the processor sees no complex field samples: it performs one multiply--accumulate per loop and sample, with no division and no per-branch coherent receiver. The digital form makes the step size, the polarity, the reset, and the acquisition sequence of Section~\ref{subsec:acq} programmable, at the cost of an ADC--DAC pair per loop.

\emph{Fading and the design rule.} Because $K_{\rm pd}$ is proportional to the product of the tap and reference field amplitudes, as in~\eqref{eq:geom}, $f_c$ follows the received power: a gain fixed at the maximum working power gives a bandwidth that falls in proportion to the power when the branch fades. The design point at full power is a tradeoff between tracking and noise. For a phase disturbance with a random-walk component of diffusion coefficient $D$ (rad$/\sqrt{\rm s}$), and a readout whose noise, referred to the phase, has the one-sided spectral density $S_\varphi=(i_n/I_g)^{2}$---with $i_n$ the detector noise current density and $I_g$ the gradient current per radian at lock---the residual error of the first-order loop is
\begin{equation}
\sigma_\varphi^{2}=\frac{D^{2}}{4\pi f_c}+\frac{\pi}{2}\,S_\varphi f_c,
\label{eq:design}
\end{equation}
whose first term falls and second term rises with the bandwidth. The minimum,
\begin{equation}
f_c^{\rm opt}=\frac{D}{\pi\sqrt{2S_\varphi}},\qquad
\sigma_{\varphi,\min}^{2}=D\sqrt{S_\varphi/2},
\label{eq:designopt}
\end{equation}
is the one-time design rule that replaces the runtime normalization of the digital formulation. A phase disturbance with spectral components of amplitude $A_j$ at frequencies $f_j$ adds $\sum_j(A_j^{2}/2)\,f_j^{2}/(f_j^{2}+f_c^{2})$ to~\eqref{eq:design}---the error that a first-order loop leaves on each component---which reduces to $(\Omega/2\pi f_c)^{2}$ when all components lie well inside the bandwidth, with $\Omega$ the rms rate of change of the phase, and moves the optimum up. The polarization loop follows the same rule with its disturbance and noise referred to $\theta_i$. The optimum is shallow---doubling or halving $f_c$ raises $\sigma_\varphi^{2}$ by only $25\%$---so a bandwidth that changes with the received power stays near it over a range of fades, and gain headroom keeps the loop fast enough in deeper ones (Section~\ref{sec:limits}).

\subsection{Self-Reference Tap and Residual}
\label{subsec:residual}
The reference of every readout unit comes from the monitoring tap of the combiner output, attenuated and split one-to-$N$ so that it reaches each unit at the same nominal level as the branch tap---for aligned branches, a field-amplitude ratio of $1/N$ relative to the combined field---which maximizes the contrast of the balanced interference. This is a static level setting fixed at calibration, not a per-branch amplitude control, and the resulting constant factor is absorbed into the loop gain. Because the reference and the branch tap derive from the same source and carry the same carrier and modulation, averaging over the data waveform keeps the correlation term, while the noise terms of independent branches average down.

The residual energy $\norm{\mathbf{s}/N-\mathbf{v}_i}^{2}$, the deviation of the branch from the array mean, can serve as a branch-health indicator. It is read at one port of an additional coupler set to the in-phase component ($\gamma=\pi/2$, no rotator), or reconstructed electrically from the in-phase readout of~\eqref{eq:gammaselect} and two power taps as $\norm{\mathbf{s}}^{2}/N^{2}+\norm{\mathbf{v}_i}^{2}-(2/N)\Real\{\mathbf{s}^{H}\mathbf{v}_i\}$, at the cost of extra photodiodes. It is not, however, an unbiased estimate of the branch noise: for aligned branches $\mathbf{v}_i=\mathbf{x}+\mathbf{n}_i$ with independent noise $\E\norm{\mathbf{n}_j}^{2}=\sigma_j^{2}$,
\begin{equation}
\E\Big\|\frac{1}{N}\mathbf{s}-\mathbf{v}_i\Big\|^{2}
=\Big(1-\frac{1}{N}\Big)^{2}\sigma_i^{2}
+\frac{1}{N^{2}}\sum_{j\neq i}\sigma_j^{2},
\label{eq:bias}
\end{equation}
which reduces to $(1-1/N)\sigma^{2}$ for equal branch noise. The residual is therefore nonzero even for a perfectly aligned branch and, at small $N$, carries a share of the noise of every other branch; no amplitude control is derived from it in the present scheme.

\subsection{Acquisition, Loss of Lock, and Scaling to Large Arrays}
\label{subsec:acq}
At startup only the phase loops are closed; the polarization loops are enabled once the phase loops have locked, because by~\eqref{eq:geom} the polarization gradient is weighted by $\cos\Delta\psi_i$ and points the wrong way while a branch is more than $90^{\circ}$ out of phase. On loss of lock the polarization loops freeze and the phase loops reacquire alone.

The one-to-$N$ reference distribution keeps the reference of every unit at the level of its branch tap for aligned branches, but its excess loss, like that of the combiner, grows with $\log_2N$. The front-end EDFA of every branch, ahead of the actuators and taps, sets the power at all readout units: with enough gain the readout is limited by signal--ASE beating rather than by detector noise, so that passive losses in the distribution consume power budget rather than signal-to-noise ratio (Section~\ref{subsec:simmodel}), and the same gain serves the coherent receiver. Because the readout units are independent, the resources that grow with $N$ are optical power and detectors, not an orthogonal signalling dimension such as a dither frequency, a code, or a time slot per branch. For large $N$ the readout units can alternatively be time-, wavelength-, or switch-multiplexed onto fewer detectors, trading measurement throughput against hardware count, and the combiner can be a cascade of $2\times2$ stages with local self-reference loops per stage.

Per branch, the control path comprises one PM, one PC, one readout unit (two 3~dB couplers, a Faraday rotator, two static offsets, and two balanced detectors), and two low-rate loops whose bandwidth has to follow the turbulence and pointing dynamics rather than the symbol rate; for phase rates of up to about one radian per atmospheric time constant, with time constants above 1~ms~\cite{Larsson2022}, kilohertz to tens of kilohertz suffice. The data path requires exactly one dual-polarization coherent receiver and its digital signal processing regardless of $N$, whereas per-aperture detection~\cite{Geisler2016,Liu2023} requires $N$ coherent receivers, $N$ ADC pairs, and the associated processing.

\section{Simulated Comparison with Perturbative Controllers}
\label{sec:compare}
No experiment accompanies this work, so the scheme is compared numerically with three established controllers of optical combiners: centralized and distributed stochastic parallel gradient descent (SPGD)~\cite{Zhou2009,Yang2017} and dither locking~\cite{Larsson2022}. The comparison is confined to the phase. The reference controllers act on the phase only---their scalar power and dark-port observations carry no polarization information---so the fields are taken co-polarized, the rotation $\mathbf{R}(\theta_i)$ of~\eqref{eq:model} and the polarization loops are omitted, and the proposed scheme runs its phase loops alone, in the digital form~\eqref{eq:dspphi}. Every scheme is retuned for every condition, so that control principles are compared at their best rather than at published operating points.

\begin{figure*}[t]
\centering\includegraphics[width=\textwidth]{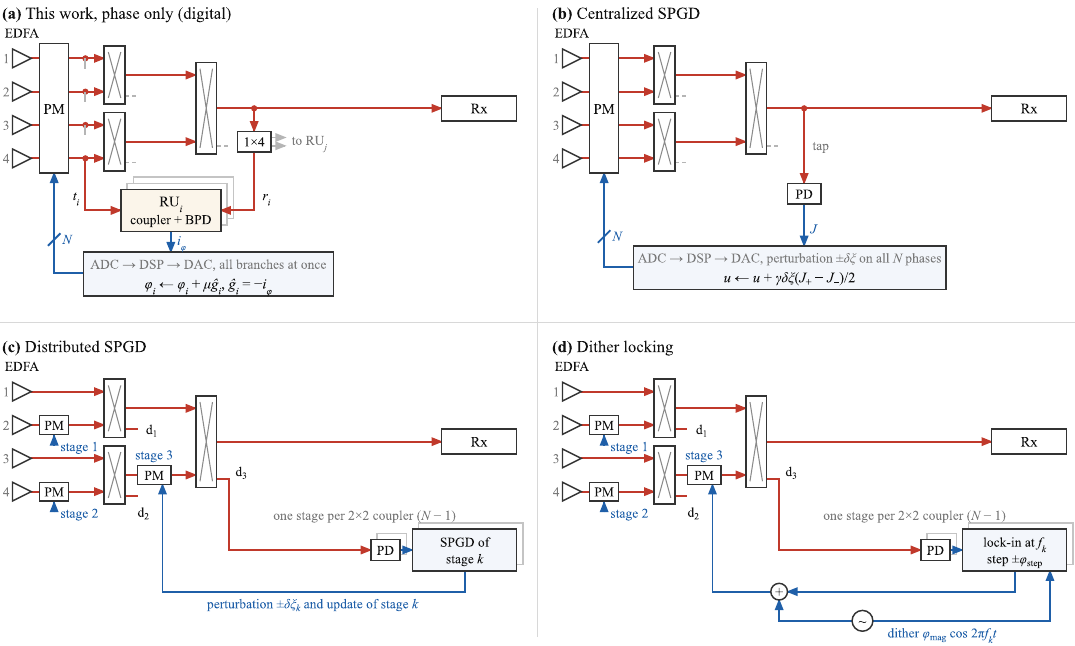}
\caption{The four phase controllers of the comparison, for $N=4$ beams summed by a binary tree of $2\times2$ couplers (red: optical; blue: electrical control). Every beam passes the same front-end EDFA (triangles), the only optical amplifier in all four schemes. (a)~This work, phase loops only, in the digital form~\eqref{eq:dspphi}: a branch tap $\mathbf{t}_i$ and a self-reference $\mathbf{r}_i$, split $1\times N$ from the monitoring tap, feed one readout unit per beam (stacked boxes), and all $N$ phases are stepped at once along the measured gradients. (b)~Centralized SPGD: one photodiode on the monitoring tap measures the combined power $J$, and the processor perturbs all $N$ phases by $\pm\delta\xi$ and updates them from $J_{+}-J_{-}$. (c)~Distributed SPGD: a photodiode detects the dark port $d_k$ of every $2\times2$ stage (stacked boxes: one detection stage per coupler), and an SPGD controller per stage drives the phase shifter in one of its input arms. (d)~Dither locking: the same detection stages; each stage dithers its phase shifter at its own frequency $f_k$, demodulates its dark-port signal, and steps by $\pm\phi_{\rm step}$. Dashed gray stubs: unused dark ports.}
\label{fig:cmpschemes}
\end{figure*}

\subsection{Model, Controllers, and Tuning}
\label{subsec:simmodel}
The $N$ beams have unit amplitude and phases $\phi_i(t)+u_i(t)$, where $\phi_i(t)$ is the phase disturbance of beam~$i$ and $u_i(t)$ is the compensating phase that the controller applies through the phase actuator of that beam; an ideal binary tree of $2\times2$ couplers forms the combined field and the $N-1$ dark-port fields, and the combining efficiency, the combined power relative to that of perfectly aligned beams, is
\begin{equation}
\eta(t)=\frac{1}{N^{2}}\Big|\sum_{i=1}^{N}e^{j[\phi_i(t)+u_i(t)]}\Big|^{2}.
\label{eq:eta}
\end{equation}
The disturbance follows the multi-tone model of the digital-combining study~\cite{Chen2026a}:
\begin{equation}
\phi_i(t)=\sum_{j=1}^{64}A_0\,s_{ij}\sin(2\pi f_jt+\psi_{ij}),\qquad f_j=\frac{j}{64}f_{\max},
\label{eq:tones}
\end{equation}
with $A_0=A\sqrt{2/64}$, random signs $s_{ij}=\pm1$, and phases $\psi_{ij}$ uniform in $[0,2\pi)$, drawn independently for every beam and realization. The rms phase $A=5$~rad is the nominal value of~\cite{Chen2026a}; it also leaves the phases at loop closure uniformly random. The rms rate of change of the phase, $\Omega=2\pi Af_{\max}\sqrt{65\cdot129/6}\,/64\approx18.4\,f_{\max}$, is swept from 10 to $10^{5}$~rad/s by $f_{\max}=0.54$~Hz--5.4~kHz, and up to $10^{7}$~rad/s ($f_{\max}=545$~kHz) in Section~\ref{subsec:simber}. Atmospheric time constants above 1~ms bound turbulence to $\Omega\lesssim10^{3}$~rad/s~\cite{Larsson2022}, and the inter-channel rate measured there is 51~rad/s; the higher rates test the margin of every controller, as the synthetic disturbance of~\cite{Chen2026a} does for digital combiners.

Every beam passes the same EDFA directly after its aperture (noise figure 5~dB), the only optical amplifier in all four schemes. The nominal operating point is the OSNR at which ideal combining of four beams just reaches the hard-decision FEC threshold of the DP-QPSK signal of Section~\ref{subsec:simber}: a received power of $-47$~dBm per beam, amplified by 47~dB to an output of 0~dBm, gives an OSNR per beam---the signal power over the ASE power in both polarizations within $B_{\rm ref}=12.5$~GHz (0.1~nm)---of 6.0~dB. Every study runs at this point unless it varies the OSNR or the power. Branch loss of 4~dB and excess loss of 0.35~dB per coupler follow. Each detector output, normalized to the full-scale signal, carries white Gaussian noise whose variance follows from the thermal ($18$~pA$/\sqrt{\rm Hz}$), shot, signal--ASE, and ASE--ASE noise at that detector. In the proposed scheme, 5\% branch and monitoring taps deliver $-17$ and $-18$~dBm to the readout coupler, a gradient current of 35.8~$\mu$A/rad, and a readout noise of 3.8~$\mu$rad$/\sqrt{\rm Hz}$ referred to the phase, set mainly by signal--ASE beating; with a front-end gain of only 20~dB, the thermal noise would exceed it by 35~dB. Centralized SPGD reads the same 5\% monitoring tap with one photodiode, and distributed SPGD and dither locking read the dark port of every coupler with a photodiode, without the per-stage EDFA, filter, and polarizer of~\cite{Larsson2022}: behind the common front-end gain, the dark-port photocurrent is already limited by ASE beating rather than by thermal noise.

Fig.~\ref{fig:cmpschemes} shows the four controllers as simulated, for $N=4$. To compare control principles rather than devices, the detectors have unlimited bandwidth and all four controllers are digital: each samples its detector at $f_s=100$~MS/s, as current converters and processors allow, applies its new setting one sample later (10~ns), and is neither quantized nor range-limited. The clock is chosen so that it does not limit the proposed loop up to $10^{5}$~rad/s: its steady-state optimum stays at $\mu K_{\rm pd}\le0.19$ ($f_c\le3.1$~MHz), and the steps it selects for fast acquisition at $\mu K_{\rm pd}\le0.54$, far inside its stability range; only the extension to $10^{7}$~rad/s in Section~\ref{subsec:simber} reaches the limits of the clock. The proposed loop applies~\eqref{eq:dspphi} to every beam, $u_i\leftarrow u_i+\mu\hat{g}_i$, and its only tuning parameter is the step per sample $\mu K_{\rm pd}=2\pi f_cT_s$. Centralized SPGD~\cite{Zhou2009,Yang2017} perturbs all $N$ phases by Bernoulli steps $\pm\delta\xi_i$ in every iteration, averages the monitoring signal over each half-period except its first, delayed sample to obtain $J_{\pm}$, and updates $\mathbf{u}\leftarrow\mathbf{u}+\gamma\delta\boldsymbol{\xi}(J_{+}-J_{-})/2$; its tuning parameters are the iteration rate $f_{\rm it}$, $\delta$, and the effective gradient step $\kappa=\gamma\delta^{2}$. Distributed SPGD~\cite{Yang2017} runs the same algorithm at every $2\times2$ stage and minimizes the power of its dark port by moving one of the two input groups. Dither locking~\cite{Larsson2022} dithers one input group of every stage at $f_d$ with amplitude $\phi_{\rm mag}$, eight samples per period, demodulates the dark-port signal once per period, sums the last $m$ demodulated values, and steps the phase by $\phi_{\rm step}$ against the sign of the sum. Its frequencies are those of~\cite{Larsson2022}, one per stage (20, 15, and 25~kHz), in the rate study and one per tree level ($20,25,\dots,15+5\log_2N$~kHz) in the scaling study, all multiplied by a tuned factor that keeps the highest one at or below $f_s/8$.

Each scheme is tuned for every condition by an exhaustive search over the grid of Table~\ref{tab:grids} that maximizes the mean efficiency over the first 1~ms after loop closure on two training realizations,
\begin{equation}
J=\frac{1}{2K}\sum_{m=1}^{2}\sum_{k=1}^{K}\eta_m(t_k),\qquad 0\le t_k-t_c<1~\text{ms},
\label{eq:tunemetric}
\end{equation}
a metric that rewards fast acquisition and steady tracking alike and is the same in all studies. Here $\eta_m(t_k)$ is~\eqref{eq:eta} in realization $m$ at the record instants $t_k$, which are spaced by 101 samples (1.01~$\mu$s) so that they do not alias the periodic dither and SPGD perturbations, and $t_c$ is the instant of loop closure. The training realizations are common to all grid points, so that the points are compared on identical disturbances and noise. The grids extend every rate parameter far below the limit the clock sets---the bandwidth of the proposed loop down to 1~kHz, the SPGD iteration rate down to 0.25~MHz, and the dither frequencies down to a fiftieth of their maximum---and the optima lie inside the grids or at a limit that the clock or the algorithm imposes---the SPGD iteration rate of 25~MHz (two samples per half-period), the dither frequency $f_s/8=12.5$~MHz, and demodulation over a single dither period. The selected settings are then evaluated on $M$ independent realizations---$M=100$ in the rate study and 40 in the others---recorded for 10~ms after loop closure and once more for 0.1~ms at full time resolution. The loops are closed at $t_c=20~\mu$s, from random phases. All figures use the following metrics. The residual loss is
\begin{equation}
L=-10\log_{10}\langle\eta\rangle,\qquad \langle\eta\rangle=\frac{1}{MK}\sum_{m=1}^{M}\sum_{k=1}^{K}\eta_m(t_k),
\label{eq:resloss}
\end{equation}
over the $K$ record instants of the steady state, 1--10~ms after loop closure; the steady-state loss of Fig.~\ref{fig:cmpmu}(a) is the same quantity over 0.25--1~ms on the training realizations. The first-percentile loss $L_{1\%}=-10\log_{10}\eta_{1\%}$, with $\eta_{1\%}$ the first percentile of all $MK$ steady-state samples $\eta_m(t_k)$, captures brief, deep misalignments that the mean hides. The instantaneous loss $\ell_m(t)=-10\log_{10}\eta_m(t)$, recorded at every sample of the 0.1-ms record, shows the acquisition, and the time to $\eta=0.9$ ($\ell=0.46$~dB),
\begin{equation}
t_{90,m}=\min\{t-t_c:\ t\ge t_c,\ \bar\eta_m(t)\ge0.9\},
\label{eq:t90}
\end{equation}
with $\bar\eta_m$ the 0.5-$\mu$s trailing moving average of $\eta_m$, measures it; it is taken from the 10-ms record, without averaging, where $\bar\eta_m$ stays below 0.9 during the 0.1-ms record, and the figures give its median over the realizations.

\begin{table}[t]
\caption{Tuning Grids (Identical for Every Condition)}
\label{tab:grids}
\centering\footnotesize
\setlength{\tabcolsep}{2.5pt}
\begin{tabular}{lll}
\hline
Scheme & Parameter & Values \\
\hline
This work & $\mu K_{\rm pd}=2\pi f_cT_s$ & $1.536\cdot2^{-k/2}$, $k=0,\dots,29$ \\
 & & ($f_c=1.0$~kHz--$24$~MHz) \\
SPGD (both) & $f_{\rm it}$ & 0.25, 1, 2.5, 5, 10, 25~MHz \\
 & $\delta$ & 0.005--1~rad (15 values) \\
 & $\kappa=\gamma\delta^{2}$ & 0.003--1000 (22 values) \\
Dither locking & $\phi_{\rm step}$ & $0.2^{\circ}$--$32.8^{\circ}$ (13 values) \\
 & $\phi_{\rm mag}$ & $0.24^{\circ}$--$30.7^{\circ}$ (12 values) \\
 & freq.\ factor & 10, 50, 200, max ($f_{d,\max}=f_s/8$) \\
 & periods summed $m$ & 1, 3, 10, 30 \\
\hline
\end{tabular}
\end{table}

\subsection{Rate of the Phase Disturbance}
\label{subsec:simrate}
\begin{figure}[t]
\centering\includegraphics[width=\columnwidth]{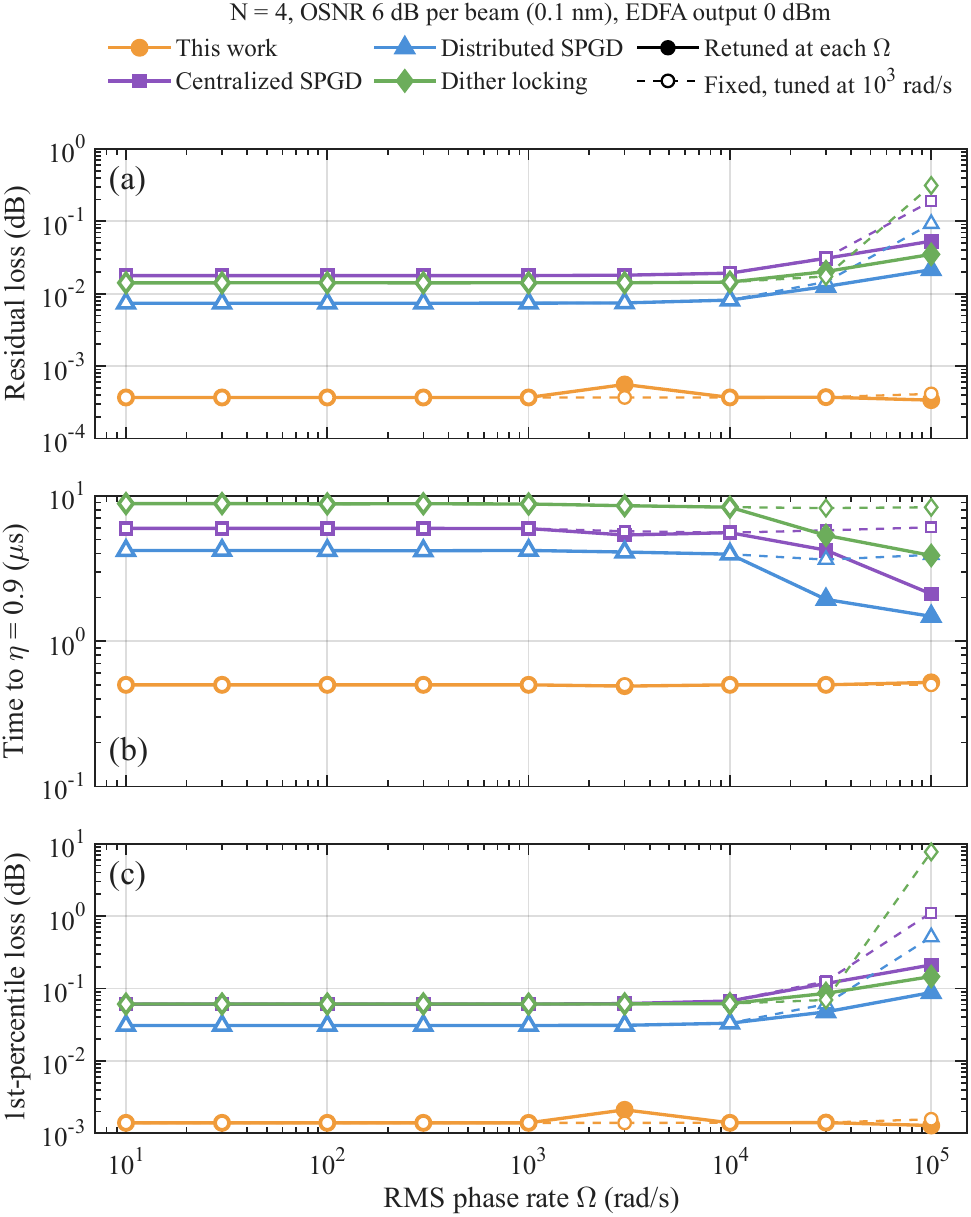}
\caption{Comparison versus the rms rate $\Omega$ of the phase disturbance~\eqref{eq:tones} ($N=4$, OSNR 6~dB per beam at an EDFA output of 0~dBm, 100 realizations, ideal devices, all controllers at 100~MS/s behind the same front-end EDFA). (a) Residual loss in the steady state; (b) median time to reach $\eta=0.9$ (0.5-$\mu$s moving average); (c) loss exceeded for 1\% of the steady-state time. Solid, filled: every scheme retuned at every rate; dashed, open: settings tuned at $10^{3}$~rad/s held fixed.}
\label{fig:cmprate}
\end{figure}

Fig.~\ref{fig:cmprate} compares the four schemes at $N=4$ and the nominal OSNR of 6~dB per beam, each retuned at every rate $\Omega$ (solid) and, for reference, with the settings tuned at $10^{3}$~rad/s held fixed (dashed); Table~\ref{tab:tuned} lists the selected settings. With the 100-MS/s clock every scheme holds the lock at every rate---none spends more than 0.15\% of the steady-state time below $\eta=0.9$---and the residual loss~\eqref{eq:resloss} reflects the measurement each relies on [Fig.~\ref{fig:cmprate}(a)]. Up to $10^{4}$~rad/s the losses hardly depend on the rate: the proposed loop loses 3.7--$5.6\times10^{-4}$~dB, distributed SPGD 0.0074--0.0082~dB, dither locking 0.014~dB, and centralized SPGD 0.018--0.019~dB, levels set by the readout noise of the proposed loop and by the perturbation, dither, and limit cycle of the baselines. Faster disturbances raise the losses of the baselines: at $10^{5}$~rad/s distributed SPGD loses 0.022~dB, dither locking 0.035~dB, and centralized SPGD 0.053~dB, whereas the proposed loop, with a slightly smaller step, loses $3.4\times10^{-4}$~dB. The proposed loop thus loses 13 to 160 times less than the baselines, and the first-percentile loss orders the schemes in the same way [Fig.~\ref{fig:cmprate}(c)]: at $10^{5}$~rad/s it is $1.3\times10^{-3}$~dB for the proposed loop and 0.088--0.21~dB for the baselines.

\begin{figure}[t]
\centering\includegraphics[width=\columnwidth]{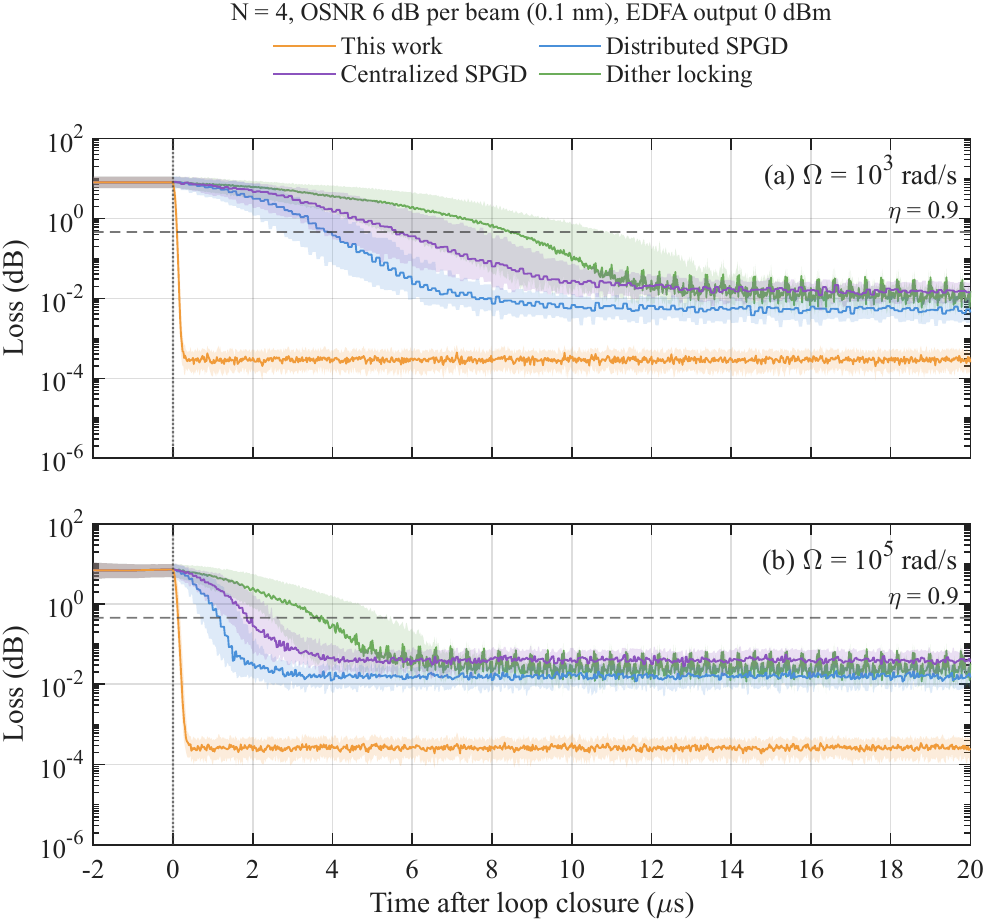}
\caption{Acquisition from random phases at (a) $\Omega=10^{3}$~rad/s and (b) $10^{5}$~rad/s, each with the settings selected for that rate ($N=4$, OSNR 6~dB per beam, Table~\ref{tab:tuned}): median instantaneous loss $\ell=-10\log_{10}\eta$ over 100 realizations (lines) and range between the 25th and 75th percentiles (shading), recorded every 10~ns, versus the time after loop closure (dotted line). Dashed: $\eta=0.9$.}
\label{fig:cmptrace}
\end{figure}

The proposed loop reaches $\eta=0.9$, by~\eqref{eq:t90}, within 0.5~$\mu$s at every rate, a time set by the 0.5-$\mu$s averaging of the metric: its median efficiency crosses 0.9 after 0.10--0.13~$\mu$s, ten to thirteen samples after loop closure (Fig.~\ref{fig:cmptrace}). Distributed SPGD needs 1.5--4.2~$\mu$s, centralized SPGD 2.1--6.0~$\mu$s, and dither locking 3.9--8.8~$\mu$s [Fig.~\ref{fig:cmprate}(b)]. The retuned acquisition times of the baselines fall at the highest rates, where the tuning selects faster settings; with the settings held fixed (dashed), every acquisition time is independent of the rate.

The selected settings show which limits remain. Dither locking selects the highest dither frequencies, $f_s/8$, and a step on every dither period at every rate, because a sign-driven controller follows at most $\phi_{\rm step}f_d$~\cite{Larsson2022} and acquires the faster the higher $f_d$; since it needs eight samples per dither period, its speed remains tied to the clock. The SPGD variants select iteration rates of 5--10~MHz at every rate, below the limit of 25~MHz, two samples per half-period, that the clock sets: the noisy power readings at this OSNR favor longer averaging over faster iteration, and at an OSNR of 8~dB and above centralized SPGD does select 25~MHz at $10^{5}$~rad/s (Table~\ref{tab:tuned}). The proposed loop selects $\mu K_{\rm pd}=0.27$ ($f_c=4.3$~MHz), 0.38 at $3\times10^{3}$~rad/s, and 0.19 at $10^{5}$~rad/s: these steps acquire within about 0.1~$\mu$s, whereas the smaller steps that would track with less noise acquire more slowly, which the metric~\eqref{eq:tunemetric} penalizes. All of them lie far inside the stability range, so the clock does not limit the loop (Section~\ref{subsec:simmu}). Held at the settings tuned at $10^{3}$~rad/s [Fig.~\ref{fig:cmprate}, dashed], the proposed loop loses $4.1\times10^{-4}$~dB at $10^{5}$~rad/s: the error $(\Omega/2\pi f_c)^{2}$ that its 4.3-MHz bandwidth leaves on the disturbance adds $4.5\times10^{-5}$~dB to its readout noise. Distributed SPGD, centralized SPGD, and dither locking lose 0.093, 0.19, and 0.31~dB there and spend 1.4\%, 8.8\%, and 15\% of the time below $\eta=0.9$.

Fig.~\ref{fig:cmptrace} shows the acquisition from random phases at $10^{3}$ and $10^{5}$~rad/s, each with the settings selected for that rate; the instantaneous loss is plotted on a logarithmic scale because every steady state lies within 0.05~dB of perfect alignment. The median efficiency of the proposed loop crosses 0.9 after 0.10 and 0.13~$\mu$s. With the faster settings selected at $10^{5}$~rad/s, the baselines acquire faster there than at $10^{3}$~rad/s: distributed SPGD after 1.2 instead of 3.9~$\mu$s, centralized SPGD after 1.9 instead of 5.6~$\mu$s, and dither locking after 3.5 instead of 8.4~$\mu$s. The acquisition thus takes a few microseconds, during which a disturbance of $10^{5}$~rad/s moves the phases by at most about 0.4~rad~rms. The faster disturbance shows in the steady state: the median loss rises from 0.0056 to 0.015~dB for distributed SPGD, from 0.012 to 0.026~dB for dither locking, and from 0.015 to 0.039~dB for centralized SPGD, whereas that of the proposed loop stays at 2.5--$2.9\times10^{-4}$~dB.

\subsection{Step Size of the Digital Loop}
\label{subsec:simmu}
\begin{figure}[t]
\centering\includegraphics[width=\columnwidth]{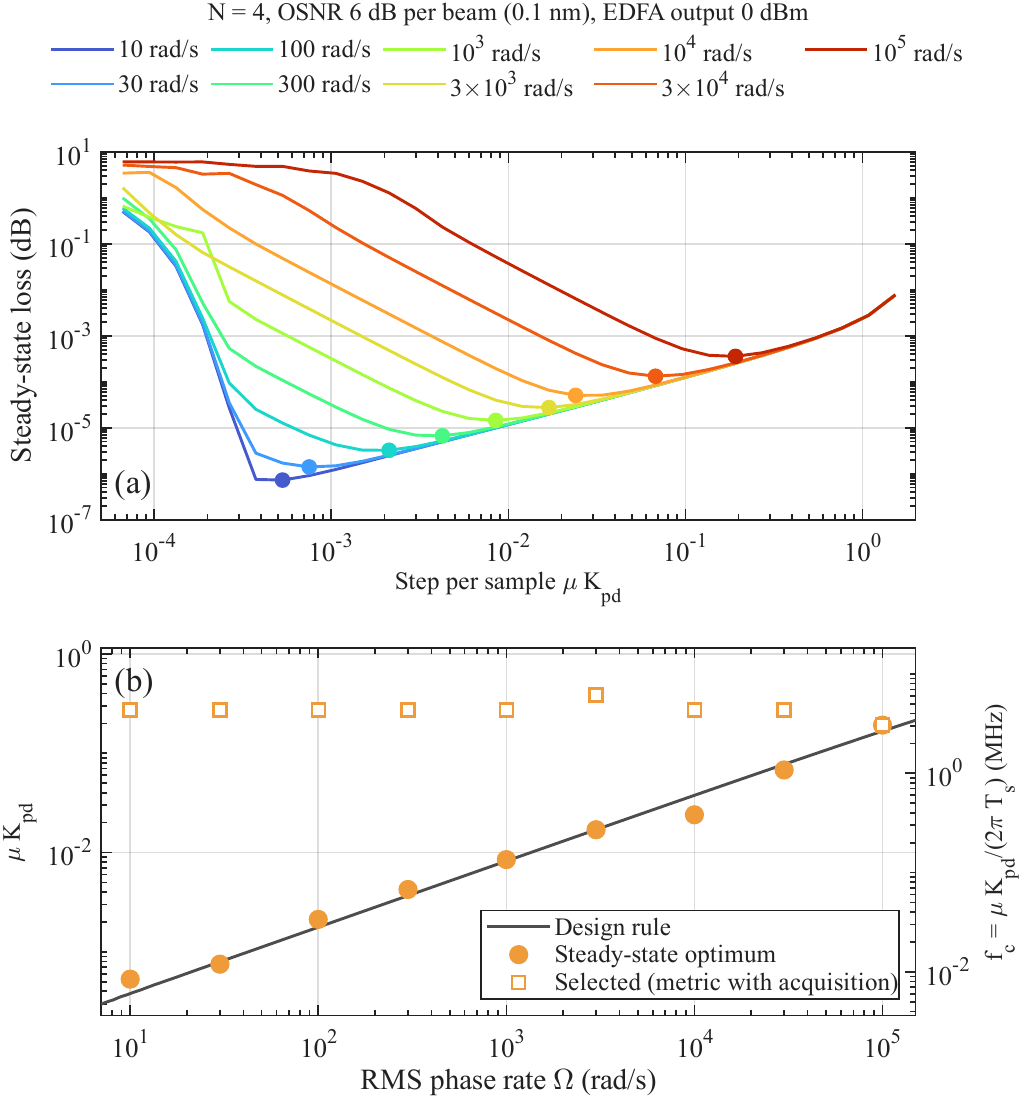}
\caption{Step size of the digital loop~\eqref{eq:dspphi} ($N=4$, OSNR 6~dB per beam, $f_s=100$~MS/s). (a) Steady-state loss (mean over 0.25--1~ms after loop closure, training realizations) versus the step per sample $\mu K_{\rm pd}=2\pi f_cT_s$ for each rate $\Omega$; dots: minima. (b) Steady-state optimum (filled), step selected with the metric that includes acquisition (open), and the optimum of the design rule in its sampled form (line); right axis: the corresponding bandwidth $f_c$.}
\label{fig:cmpmu}
\end{figure}

Fig.~\ref{fig:cmpmu}(a) shows the steady-state loss of the proposed loop---the mean over 0.25--1~ms after loop closure on the training realizations---versus its step per sample $\mu K_{\rm pd}$. Small steps lag the disturbance and large steps pass the readout noise, so that at large steps the curves merge into one noise-limited branch. Every curve has its minimum inside the grid, from $\mu K_{\rm pd}=5.3\times10^{-4}$ ($f_c=8.4$~kHz, loss $7.3\times10^{-7}$~dB) at 10~rad/s to 0.19 ($f_c=3.1$~MHz, loss $3.6\times10^{-4}$~dB) at $10^{5}$~rad/s: with the 100-MS/s clock even the fastest disturbance calls for a step far inside the stability range $\mu K_{\rm pd}<2$. Steps below about $3\times10^{-4}$ ($f_c\lesssim5$~kHz) have not completed their acquisition when the window opens at 0.25~ms, which raises the left end of every curve steeply. For the sampled loop, the design rule~\eqref{eq:design} takes its exact form: the noise term becomes $\sigma_g^{2}\,\mu K_{\rm pd}/(2-\mu K_{\rm pd})$, with $\sigma_g^{2}=S_\varphi/(2T_s)$ the readout noise per sample, and the error the loop leaves on a tone at $f_j$ becomes $(A_j^{2}/2)\,|H_e(e^{j2\pi f_jT_s})|^{2}$, with $H_e(z)=(z-1)/(z-1+\mu K_{\rm pd})$; both reduce to the terms of Section~\ref{subsec:loopimpl} for $\mu K_{\rm pd}\ll1$. With the tones~\eqref{eq:tones} and the $S_\varphi$ of the simulated link, the rule predicts the steady-state optimum within about one grid step at every rate [Fig.~\ref{fig:cmpmu}(b)]: $3.8\times10^{-4}$, 0.0082, and 0.17 at 10, $10^{3}$, and $10^{5}$~rad/s, against $5.3\times10^{-4}$, 0.0085, and 0.19; at $10^{4}$~rad/s, where it predicts 0.038 against 0.024, the curve is flat to 1\% between 0.024 and 0.034. Because all tones lie far inside the bandwidth, the rule is close to the closed form $f_c^{\rm opt}=(\Omega^{2}/\pi^{3}S_\varphi)^{1/3}$, the minimum of $(\Omega/2\pi f_c)^{2}+(\pi/2)S_\varphi f_c$, which agrees within 6\% up to $10^{5}$~rad/s: the optimal bandwidth grows as $\Omega^{2/3}$. The step selected for the comparison is larger at low rates, 0.27--0.38 up to $3\times10^{4}$~rad/s, because the metric that includes acquisition favors fast loops and the readout noise costs little: at 10~rad/s it raises the steady-state loss from $7.3\times10^{-7}$ to $3.7\times10^{-4}$~dB. The single parameter of the proposed loop can therefore be set from the link budget and the expected phase dynamics without a search, and raised where fast acquisition matters.

\subsection{Number of Beams}
\label{subsec:simn}
\begin{figure}[t]
\centering\includegraphics[width=\columnwidth]{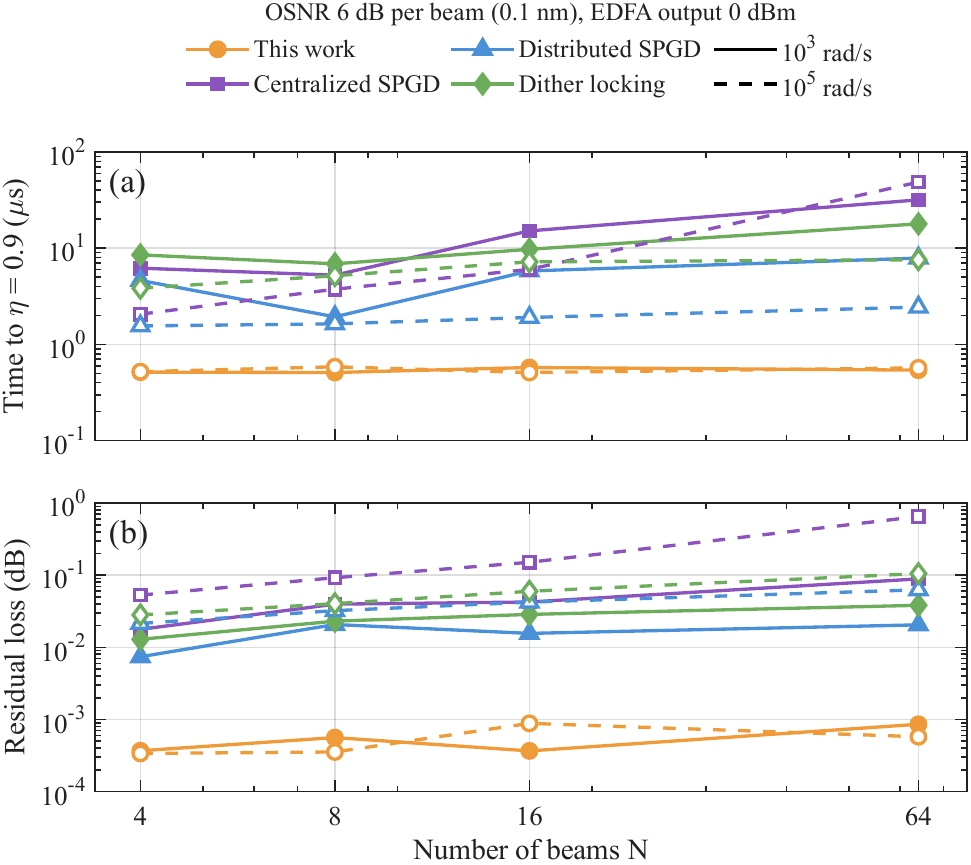}
\caption{Scaling with the number of beams (OSNR 6~dB per beam, 40 realizations, every scheme retuned for every $N$ and rate, 100~MS/s): (a) median time to reach $\eta=0.9$ (0.5-$\mu$s moving average); (b) residual loss. Solid, filled: $\Omega=10^{3}$~rad/s; dashed, open: $10^{5}$~rad/s.}
\label{fig:cmpscaling}
\end{figure}

Fig.~\ref{fig:cmpscaling} extends the comparison to $N=8$, 16, and 64 at $10^{3}$ and $10^{5}$~rad/s and the nominal OSNR of 6~dB, with every scheme retuned for every $N$ and rate. Because all $N$ gradients are read at once, the proposed loop reaches $\eta=0.9$ within 0.51--0.59~$\mu$s at every $N$ and rate [Fig.~\ref{fig:cmpscaling}(a)], and its residual loss stays between $3.4\times10^{-4}$ and $8.8\times10^{-4}$~dB [Fig.~\ref{fig:cmpscaling}(b)], as expected, because the efficiency penalty of independent phase errors of variance $\sigma_\phi^{2}$, $1-\langle\eta\rangle\simeq(1-1/N)\sigma_\phi^{2}$, saturates with $N$; most of the remaining variation follows the step that the tuning selects, from $\mu K_{\rm pd}=0.19$ to 0.54. Distributed SPGD, whose stages also work in parallel, acquires within 1.6--7.9~$\mu$s at every $N$, but its loss grows from 0.0074 to 0.021~dB at $10^{3}$~rad/s and from 0.021 to 0.063~dB at $10^{5}$~rad/s, because every beam passes $\log_2N$ stages whose residual errors add. Centralized SPGD, which extracts $N$ directions from one scalar, slows from 2.1--6.2~$\mu$s at $N=4$ to 32~$\mu$s at $N=64$ and $10^{3}$~rad/s, where it loses 0.089~dB, and at $N=64$ and $10^{5}$~rad/s it no longer holds the lock: even at the iteration limit of 25~MHz it spends 91\% of the steady-state time below $\eta=0.9$ and loses 0.65~dB. Dither locking, with $\log_2N$ cascaded tree levels and frequencies capped at $f_s/8$, needs 3.8--18~$\mu$s and loses 0.013--0.11~dB. Every other scheme holds the lock at every $N$ and rate, and both SPGD variants select the iteration rate of 25~MHz at $N=64$ and $10^{5}$~rad/s.

\subsection{OSNR at Constant Power and EDFA Output Power}
\label{subsec:simosnr}
\begin{figure}[t]
\centering\includegraphics[width=\columnwidth]{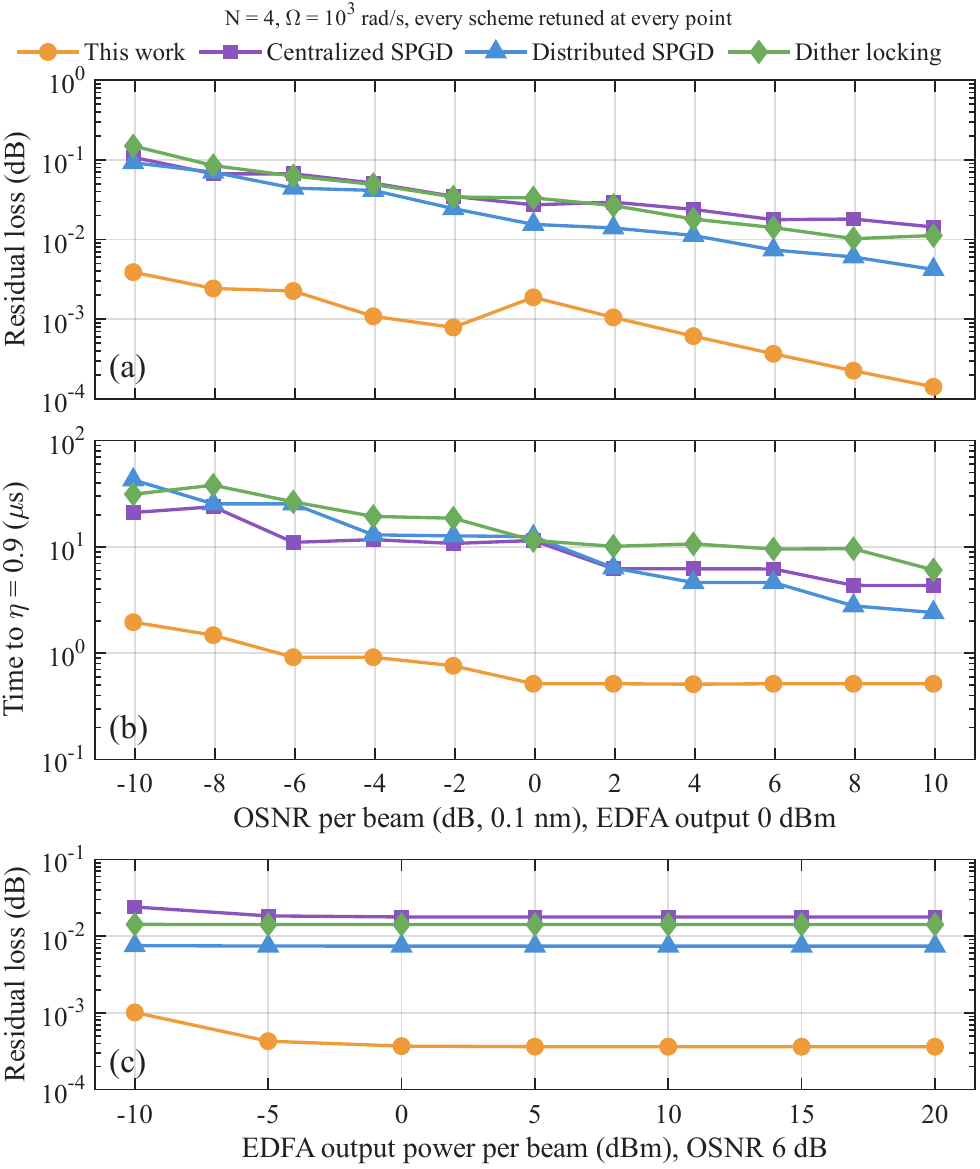}
\caption{OSNR and EDFA output power ($N=4$, $\Omega=10^{3}$~rad/s, 40 realizations, 100~MS/s, every scheme retuned at every point). (a) Residual loss~\eqref{eq:resloss} and (b) median time to $\eta=0.9$~\eqref{eq:t90} versus the OSNR per beam, varied at a constant EDFA output of 0~dBm per beam so that only the ASE changes. (c) Residual loss versus the EDFA output power per beam at the nominal OSNR of 6~dB.}
\label{fig:cmposnr}
\end{figure}

Fig.~\ref{fig:cmposnr} varies the OSNR per beam from 10 to $-10$~dB in 2-dB steps while every EDFA delivers the same signal power of 0~dBm: the received power drops from $-43$ to $-63$~dBm ahead of an EDFA whose gain rises from 43 to 63~dB, so that only the ASE grows, whereas the signal powers at all detectors, and with them the loop gains, keep their nominal values ($\Omega=10^{3}$~rad/s). At low OSNR the ASE--ASE beating dominates, and the noise per sample of every detector then grows about tenfold for every 10~dB. Every scheme is retuned at every OSNR.

Every scheme loses more as the OSNR falls [Fig.~\ref{fig:cmposnr}(a)], but the proposed loop does so at a much lower level: it loses $1.4\times10^{-4}$~dB at 10~dB, $3.7\times10^{-4}$~dB at the nominal 6~dB, and $3.9\times10^{-3}$~dB at $-10$~dB, against 0.0042--0.014, 0.0074--0.018, and 0.092--0.15~dB for the baselines, 8 to 100 times less over the sweep. The factor is smallest at 0~dB, where the loss of the proposed loop rises to $1.9\times10^{-3}$~dB before it falls again at $-2$~dB: down to 0~dB the tuning metric, which also rewards acquisition, keeps the step $\mu K_{\rm pd}=0.27$, whereas from $-2$~dB on it selects 0.068 and less, whose lower readout noise outweighs the slower acquisition. The proposed loop reaches $\eta=0.9$ within 0.5~$\mu$s down to 0~dB and within 2~$\mu$s at $-10$~dB, whereas the baselines need 2.4--9.6~$\mu$s at 10~dB and 21--43~$\mu$s at $-10$~dB [Fig.~\ref{fig:cmposnr}(b)], where dither locking already spends 3.8\% of the steady-state time below $\eta=0.9$. No scheme fails within this range of OSNR, which brackets the FEC threshold of the data path. A sweep extended to $-35$~dB with the same model and tuning (not shown) finds the baselines failing---more than half of the steady-state time below $\eta=0.9$---from about $-20$ to $-22$~dB on, and the proposed loop only at about $-32$~dB.

The margin follows from what each scheme measures. The balanced readout delivers the phase error itself, linearly and at every sample, so the loop only has to average the noise over its bandwidth, which the tuning lowers from 4.3~MHz at high OSNR to 0.27~MHz at $-10$~dB. The baselines measure powers, which near lock change only quadratically with the error; they have to probe them with perturbations or dithers, which the tuning enlarges at low OSNR---$\delta$ from 0.03--0.05 to 0.1~rad and $\phi_{\rm mag}$ from $2.9^{\circ}$ to $15^{\circ}$---at a cost in efficiency of their own.

Raising the EDFA output above 0~dBm changes nothing that matters [Fig.~\ref{fig:cmposnr}(c)]: at 0~dBm every detector is already limited by beating with the ASE, which grows with the signal, so a higher output raises signal and noise alike, and the losses at $+20$~dBm differ from those at 0~dBm by less than 2\%. A lower output lets the thermal noise in: at $-10$~dBm the proposed loop loses $1.0\times10^{-3}$~dB, 2.7 times more than at 0~dBm, and centralized SPGD 0.024~dB (35\% more), whereas distributed SPGD and dither locking change by about 1\%. The front-end gain therefore only has to lift the signal to about 0~dBm per beam; in a fade the EDFA output falls with the received power, which Section~\ref{subsec:simpower} examines.

\subsection{Fading with a Fixed-Gain Front End}
\label{subsec:simpower}
\begin{figure}[t]
\centering\includegraphics[width=\columnwidth]{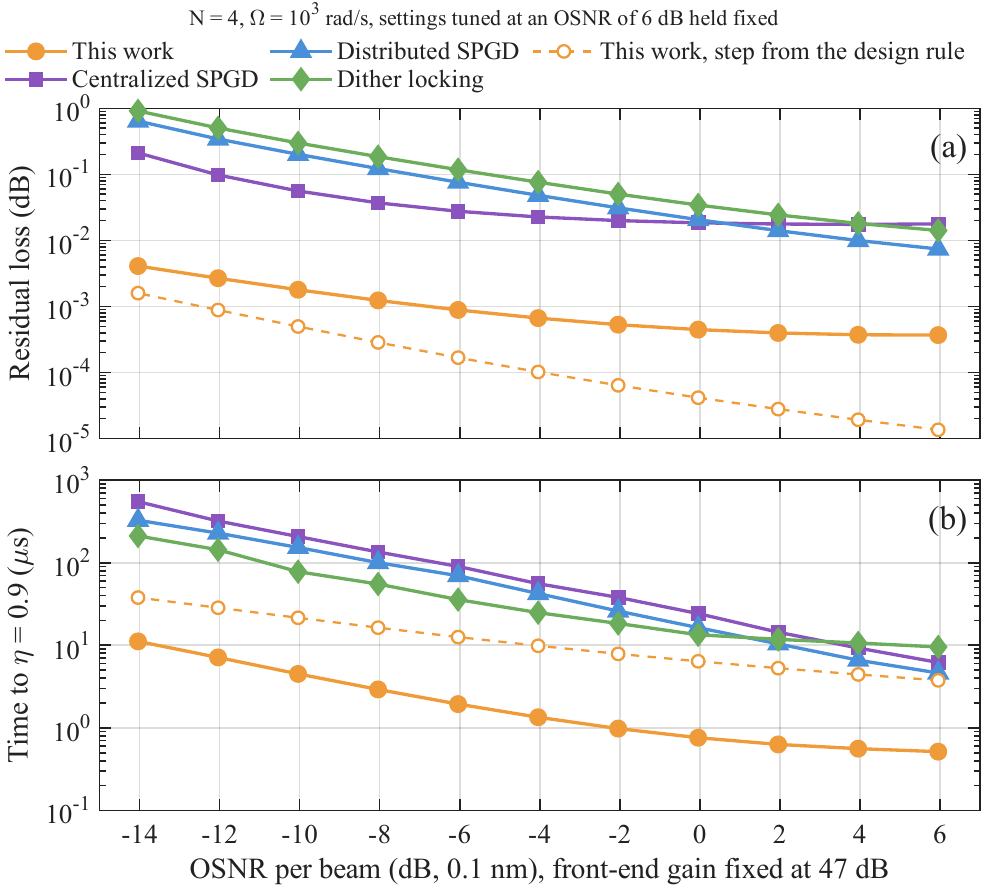}
\caption{Fading with a fixed-gain front end: (a) residual loss and (b) median time to $\eta=0.9$ versus the OSNR per beam in 0.1~nm, lowered by lowering the received power ahead of the fixed-gain front-end EDFA (OSNR in dB $=P+53$, with $P$ the received power per beam in dBm; $N=4$, $\Omega=10^{3}$~rad/s, 40 realizations, 100~MS/s). Solid, filled: every scheme keeps the gains and settings tuned at the nominal 6~dB; dashed, open: this work with its step scheduled from the sampled design rule.}
\label{fig:cmppower}
\end{figure}

Fig.~\ref{fig:cmppower} lowers the OSNR of every beam at $\Omega=10^{3}$~rad/s as a fade does: the received power drops ahead of the fixed-gain front-end EDFA, so the OSNR per beam, 6~dB at the nominal $-47$~dBm, falls by the same number of decibels, to $-14$~dB at $-67$~dBm, and, unlike in Section~\ref{subsec:simosnr}, the powers at all detectors fall with it; the OSNR is stepped in 2-dB increments. Every scheme keeps the gains and settings tuned at the nominal OSNR, as it would during a fade faster than any adaptation. A fade slow enough for retuning adds nothing to Fig.~\ref{fig:cmposnr}: every detector noise term that matters here---signal--ASE and ASE--ASE beating---scales with the signal like the gradients and powers it corrupts, so a retuned scheme at a given OSNR sees the same noise relative to its signal at either power, apart from the thermal and shot noise, which stay 8~dB or more below the ASE beating over the whole fade.

The proposed loop has the lowest loss at every OSNR [Fig.~\ref{fig:cmppower}(a)]: $3.7\times10^{-4}$~dB at 6~dB, $4.5\times10^{-4}$~dB at 0~dB, and $4.1\times10^{-3}$~dB at $-14$~dB, where its bandwidth has fallen a hundredfold, to 43~kHz, and its acquisition has slowed from 0.5 to 11~$\mu$s [Fig.~\ref{fig:cmppower}(b)]. This is the fading-dependent gain of Section~\ref{subsec:loopimpl}; it costs little here, because at $-14$~dB the bandwidth still exceeds the optimum of the design rule, 12~kHz. The two SPGD variants, whose signals also scale with the received power, degrade faster: distributed SPGD loses 0.021~dB at 0~dB and 0.64~dB at $-14$~dB, and centralized SPGD 0.019 and 0.21~dB, while their acquisition slows to 0.33 and 0.55~ms at $-14$~dB. Dither locking, whose sign decisions rest on a single dither period of 80--133~ns, degrades the most: 0.034~dB at 0~dB and 0.91~dB at $-14$~dB, where it spends 65\% of the time below $\eta=0.9$. Over the fade the baselines lose 20 to 220 times more than the proposed loop.

The dashed line shows the digital loop scheduling its step from the sampled design rule of Section~\ref{subsec:simmu} for the power it monitors---through a slow photodiode on the monitoring tap or the common-mode outputs of the balanced detectors---with the readout noise at that power. The effective step $\mu K_{\rm pd}$ then falls from 0.0082 at an OSNR of 6~dB to 0.0047 at 0~dB and $7.6\times10^{-4}$ at $-14$~dB, more slowly than the gain, so the programmed step $\mu$ rises 9.3-fold over the 20~dB. The scheduled loop loses $1.4\times10^{-5}$~dB at 6~dB, $4.2\times10^{-5}$~dB at 0~dB, and $1.6\times10^{-3}$~dB at $-14$~dB, but reaches $\eta=0.9$ only within 3.8--38~$\mu$s: the design rule minimizes the steady-state error and leaves the acquisition slower than the step selected with the acquisition-weighted metric. With the same front-end amplifier for every scheme, none of them profits from an amplified detection stage of its own, and the exact gradient readout keeps its advantage at every simulated OSNR. The simulated fades are static and common to all beams; with independent fades, each loop would schedule its step from the power of its own branch.

\begin{figure*}[t]
\centering\includegraphics[width=\textwidth]{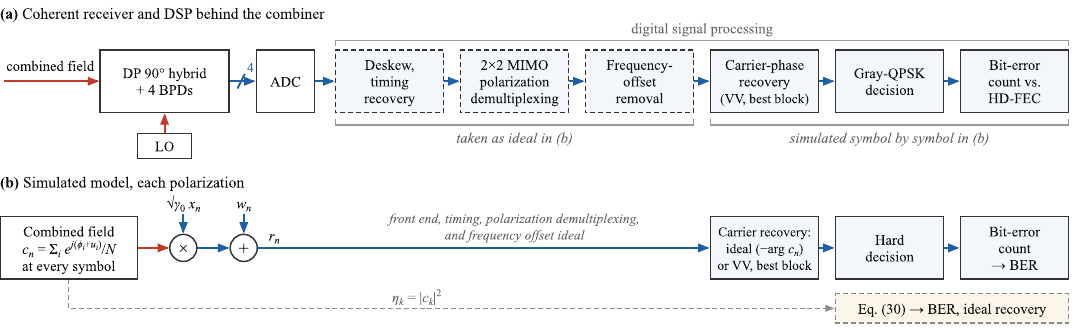}
\caption{Data path behind the combiner. (a) Coherent receiver and digital signal processing (DSP) of the 28-GBd DP-QPSK signal: the combined field is mixed with the local oscillator (LO) in a dual-polarization (DP) $90^{\circ}$ hybrid and detected by four balanced photodiodes (BPDs); behind the ADCs the DSP corrects the skew and recovers the timing, separates the polarizations with a $2\times2$ multiple-input multiple-output (MIMO) equalizer, removes the frequency offset, recovers the carrier phase---including the common phase $\arg c$ of the combined field---with a Viterbi--Viterbi (VV) block estimator whose block length is chosen for every scheme and rate, decides the Gray-coded symbols, and counts the bit errors against the HD-FEC threshold. Dashed: blocks taken as ideal in the evaluation. (b) The model evaluated for each polarization: the combined field $c_n$, formed at every symbol instant from the disturbance of every beam and the actuator settings held over each 10-ns sample, scales the QPSK symbols $x_n$ at the symbol SNR $\gamma_0$ of ideal combining, and the combined ASE of the front-end EDFAs adds the noise $w_n$; the symbol-level simulation recovers the carrier ideally ($-\arg c_n$) or blindly (VV), decides, and counts the bit errors, whereas~\eqref{eq:ber} gives the BER with ideal carrier recovery from the efficiency $\eta_k=|c_k|^{2}$ sampled every 10~ns alone.}
\label{fig:dsp}
\end{figure*}

\begin{figure}[!t]
\centering\includegraphics[width=\columnwidth]{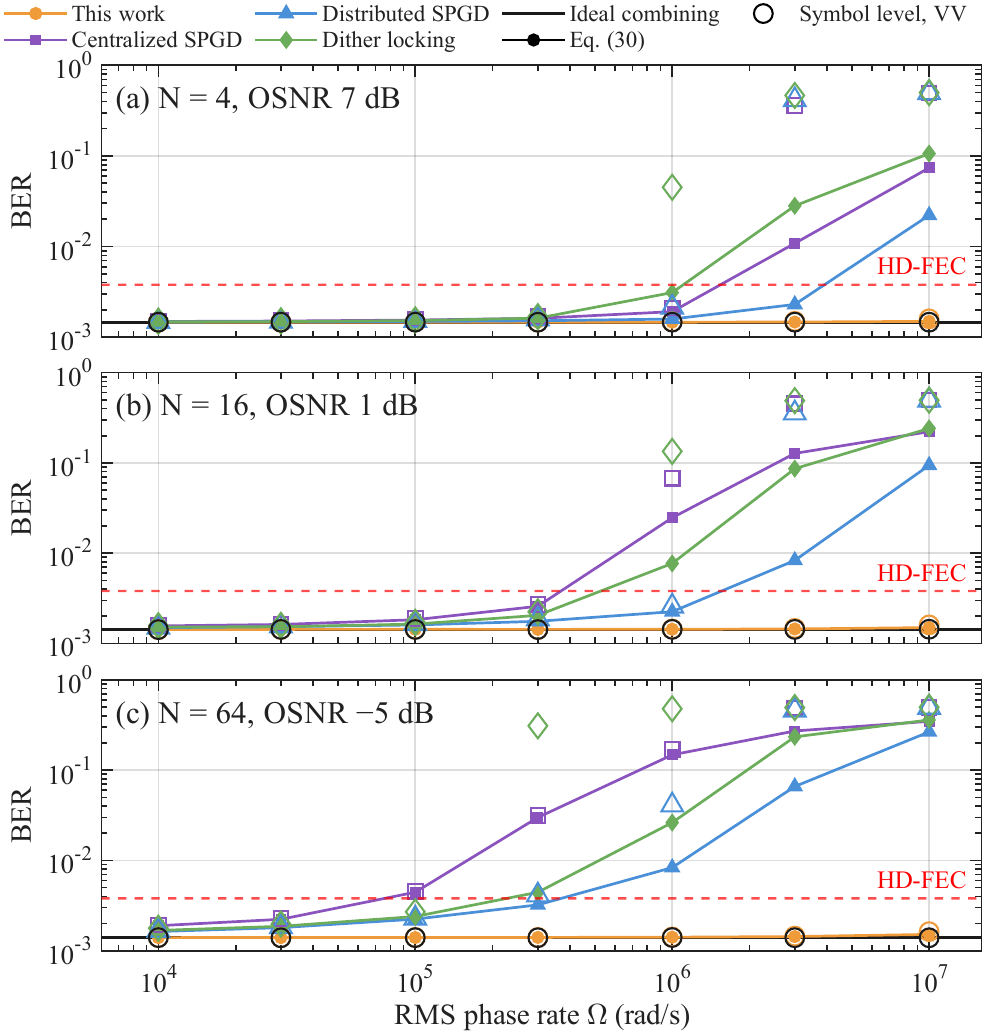}
\caption{Bit-error ratio of a 28-GBd DP-QPSK signal after the combiner versus the rms phase rate, 1~dB above the HD-FEC threshold of ideal combining: (a) $N=4$ at an OSNR of 7~dB per beam, (b) $N=16$ at 1~dB, and (c) $N=64$ at $-5$~dB (EDFA output 0~dBm, every scheme retuned at every rate, 40 realizations). Lines, filled markers: Eq.~\eqref{eq:ber}; open markers: symbol-level simulation of the complete chain with Viterbi--Viterbi carrier recovery, its block length chosen for every scheme and rate, and without differential coding; black line: ideal combining, circles: its symbol-level simulation. Red dashed: HD-FEC threshold.}
\label{fig:cmpber}
\end{figure}

\subsection{Bit-Error Ratio of a DP-QPSK Signal}
\label{subsec:simber}
To translate the residual loss into data-path performance, the transmitter is taken to send dual-polarization QPSK at $R_s=28$~GBd, and the combined field feeds a coherent receiver. The modulation does not affect the control loops: every detector of Section~\ref{subsec:simmodel} sees either the product of two fields that carry the same data or the power of a combination of such fields, so the data enter only through $|x(t)|^{2}$, whose mean the detectors take over many symbols. Behind the front-end EDFAs the receiver is limited by their ASE. The combiner adds the $N$ signal fields coherently and the $N$ independent ASE fields incoherently, so the OSNR of the combined signal is $N\eta(t)$ times the OSNR per beam, whereas the readout taps and the excess loss of the combiner attenuate signal and ASE alike and leave it unchanged. The carrier recovery of the receiver removes the common phase, which the loops change at 25~MHz at most, far slower than the symbol rate. Fig.~\ref{fig:dsp}(a) shows the data path behind the combiner: a dual-polarization coherent receiver with a local oscillator, followed by the usual digital signal processing---deskew and timing recovery, polarization demultiplexing by a $2\times2$ MIMO equalizer, removal of the frequency offset between signal and local oscillator, carrier-phase recovery, and decision of the Gray-coded symbols; neither the free-space link nor the short fibers of the receiver add dispersion that would call for compensation. The evaluation takes the front end, the timing, the polarization demultiplexing, and the frequency offset as ideal, so that each polarization reduces to the model of Fig.~\ref{fig:dsp}(b): the combined field $c_n=\sum_ie^{j[\phi_i(t_n)+u_i(t_k)]}/N$ at symbol instant $t_n$, with the actuator settings $u_i(t_k)$ of the current 10-ns sample, scales the QPSK symbols $x_n$ to the received samples $r_n=\sqrt{\gamma_0}\,c_nx_n+w_n$, with $\gamma_0=N\,\mathrm{OSNR}\,B_{\rm ref}/R_s$ and $w_n$ unit-variance complex Gaussian noise, and only the carrier recovery and the decisions remain. With Gray-coded QPSK in each polarization and ideal carrier recovery,
\begin{equation}
\mathrm{BER}=\frac{1}{MK}\sum_{m=1}^{M}\sum_{k=1}^{K}\frac{1}{2}\operatorname{erfc}\sqrt{\frac{N\eta_m(t_k)\,\mathrm{OSNR}\,B_{\rm ref}}{2R_s}},
\label{eq:ber}
\end{equation}
with OSNR the linear OSNR per beam, so that $N\eta\,\mathrm{OSNR}\,B_{\rm ref}/R_s=\gamma_0|c_k|^{2}$ is the symbol SNR $E_s/N_0$ per polarization; the average runs over the steady-state samples of~\eqref{eq:resloss}.

Versus the OSNR at constant power (Section~\ref{subsec:simosnr}), \eqref{eq:ber} shows that ideal combining of four beams reaches the HD-FEC threshold of $3.8\times10^{-3}$ at an OSNR of 6.0~dB per beam, 6.0~dB below a single aperture, and that every retuned scheme stays close to it: at $\Omega=10^{3}$~rad/s and the nominal 6~dB, the OSNR penalties, $3.7\times10^{-4}$~dB for the proposed loop and 0.0074--0.018~dB for the baselines, equal their residual losses, and at $10^{5}$~rad/s every scheme stays within 0.054~dB of ideal combining near the threshold. The BER is, however, more sensitive than the residual loss to brief, deep misalignments. At $10^{5}$~rad/s and an OSNR of 10~dB, where ideal combining gives $1.3\times10^{-5}$, the penalty of dither locking grows to 0.14~dB, seven times its residual loss of 0.020~dB, and that of centralized SPGD to 0.040~dB, whereas the penalty of the proposed loop equals its residual loss at every OSNR; with the settings tuned at $10^{3}$~rad/s held fixed [Fig.~\ref{fig:cmprate}, dashed], dither locking more than doubles its BER at $10^{5}$~rad/s and the nominal OSNR, to $9.3\times10^{-3}$, an OSNR penalty of 1.0~dB for a residual loss of 0.31~dB, and centralized and distributed SPGD pay 0.22 and 0.098~dB, whereas the proposed loop stays within $4.1\times10^{-4}$~dB. A symbol-level simulation with the combined field of every 10-ns sample held over its 280 symbols reproduces these values of~\eqref{eq:ber} with ideal carrier recovery at every OSNR from 0 to 10~dB and at both rates.

Fig.~\ref{fig:cmpber} extends the disturbance to $10^{7}$~rad/s at 1~dB above the threshold of each array size, where ideal combining gives a BER of $1.4\times10^{-3}$: four beams at an OSNR of 7~dB per beam, sixteen at 1~dB, and sixty-four at $-5$~dB, every scheme retuned at every rate (Table~\ref{tab:tuned}). This range reaches the limits of the 100-MS/s clock. From $3\times10^{5}$~rad/s on, both SPGD variants select the 25-MHz iteration rate that the clock allows, as dither locking selects its highest dither frequencies at every rate, and at $10^{7}$~rad/s the proposed loop selects steps up to $\mu K_{\rm pd}=1.54$ ($f_c=24$~MHz), close to its stability limit of 2. For four beams [Fig.~\ref{fig:cmpber}(a)], the proposed loop keeps the BER of ideal combining at every rate: its residual loss grows from $2.9\times10^{-4}$~dB at $10^{4}$~rad/s to $1.7\times10^{-3}$~dB at $10^{6}$~rad/s and 0.020~dB at $10^{7}$~rad/s, and its BER from $1.46\times10^{-3}$ to $1.50\times10^{-3}$. The baselines degrade from $10^{6}$~rad/s on, where dither locking loses 0.43~dB and spends 32\% of the time below $\eta=0.9$: by~\eqref{eq:ber}, centralized SPGD and dither locking cross the threshold at $3\times10^{6}$~rad/s, with BERs of $1.1\times10^{-2}$ and $2.8\times10^{-2}$, and at $10^{7}$~rad/s every baseline fails, with residual losses of 1.5 to 5.0~dB and BERs of $2.2\times10^{-2}$ to 0.11. More beams make the baselines fail at lower rates, whereas the proposed loop reads all $N$ gradients at every sample. For sixteen beams [Fig.~\ref{fig:cmpber}(b)], centralized SPGD and dither locking cross the threshold already at $10^{6}$~rad/s, with BERs of $2.5\times10^{-2}$ and $7.7\times10^{-3}$, and distributed SPGD at $3\times10^{6}$~rad/s, with $8.3\times10^{-3}$; for sixty-four beams [Fig.~\ref{fig:cmpber}(c)], centralized SPGD crosses it at $10^{5}$~rad/s, with $4.5\times10^{-3}$, dither locking at $3\times10^{5}$~rad/s, with $4.4\times10^{-3}$, and distributed SPGD at $10^{6}$~rad/s, with $8.3\times10^{-3}$. At $10^{7}$~rad/s their residual losses, 5.1 to 10.9~dB for sixteen beams and 12 to 17~dB for sixty-four, approach the 12 and 18~dB that these arrays lose without phase control. The proposed loop stays within 3\% of the BER of ideal combining for both array sizes up to $3\times10^{6}$~rad/s; at $10^{7}$~rad/s its residual loss, 0.043~dB for sixteen beams and 0.074~dB for sixty-four, raises the BER to $1.50\times10^{-3}$ and $1.52\times10^{-3}$. For every array size the best setting of dither locking at $10^{7}$~rad/s lies at or beyond the edge of the grid of Table~\ref{tab:grids}, but larger steps lower its loss by at most 0.5~dB.

The open markers come from a symbol-level simulation of the complete chain: random DP-QPSK symbols, the disturbance of every beam at every symbol instant, the ASE of every front-end EDFA, the setting of every actuator held over each 10-ns sample, the binary combiner, and the receiver of Fig.~\ref{fig:dsp}(a), whose Viterbi--Viterbi recovery uses, for every scheme and rate, the block length between 16 and 4096 symbols that gives the fewest errors, without differential coding (40 realizations of 0.1~ms, $4.5\times10^{8}$ bits per scheme and rate, for four beams and of 25~$\mu$s, $1.1\times10^{8}$ bits, for sixteen and sixty-four). After ideal carrier recovery its bit errors reproduce~\eqref{eq:ber} again, provided the efficiency is taken at every symbol instead of every 10~ns. The two differ because $\eta(t_k)$ is sampled right after each update, whereas the disturbance keeps drifting while the actuators hold their settings. For the proposed loop the drift adds 6\% to the loss at $10^{5}$~rad/s, also at the nominal OSNR, and less at lower rates, and for the baselines less than 1\%, so the results of Sections~\ref{subsec:simrate}--\ref{subsec:simpower} stand; at $10^{7}$~rad/s, however, it raises the loss of the proposed loop from 0.020 to 0.052~dB for four beams, from 0.043 to 0.083~dB for sixteen, and from 0.074 to 0.13~dB for sixty-four. The best block length of the proposed loop shrinks from 4096 symbols at low rates to 256--512 at $10^{7}$~rad/s, where the common phase changes faster, whereas ideal combining keeps 4096; the blind recovery then adds at most 3\% to the BER of the proposed loop with ideal carrier recovery, without a cycle slip at any rate and array size. The deep fades of the baselines, in contrast, slip cycles whatever the block length---dither locking from $10^{6}$~rad/s on for four and sixteen beams and from $3\times10^{5}$~rad/s on for sixty-four, the SPGD variants from $10^{6}$~rad/s on for sixteen and sixty-four beams---and without differential coding every slip corrupts the rest of the record: at $10^{6}$~rad/s the BER of dither locking rises to $4.5\times10^{-2}$, 0.13, and 0.48 for four, sixteen, and sixty-four beams, and from $3\times10^{6}$~rad/s on the slips nearly randomize the decisions of every baseline (BER 0.36 to 0.5). With its best block length, only SPGD with four beams avoids slips at $10^{6}$~rad/s and stays below the threshold, at $2.1\times10^{-3}$.

\subsection{Hardware and Resources}
\label{subsec:hw}
Table~\ref{tab:hw} counts the control-path hardware for phase control and its growth with $N$; the combiner, the coherent receiver, and the front-end EDFA of every beam are common to all schemes and are not counted. The schemes trade hardware against the information each measurement carries. Centralized SPGD needs one photodiode and one ADC channel regardless of $N$, but it extracts $N$ update directions from one scalar, so its acquisition slows as $N$ grows. Distributed SPGD and dither locking read the dark ports, which cost no signal power, with one photodiode per $2\times2$ stage; both perturb the beams, and dither locking needs $\log_2N$ dither frequencies on top. The proposed scheme needs the most photodiodes, $2N$ (one balanced detector per beam), and $N$ passive readout couplers with a $1\times N$ reference splitter, but no perturbation or dither generator; in its analog form it needs no ADC or DAC either. Its price is the readout taps: with 5\% branch and monitoring taps it diverts 0.45~dB of the combined power, and 0.09~dB with 1\% taps, more than the residual loss of any retuned scheme for four beams at the nominal OSNR. Behind the front-end EDFAs, however, the taps attenuate signal and ASE alike and leave the OSNR at the receiver, and with it the BER, unchanged, whereas the residual misalignment of every scheme lowers it (Section~\ref{subsec:simber}); the taps cost combined power, but no SNR, as long as the receiver is limited by the ASE. Controlling the polarization as well, which none of the reference schemes does, adds $2N$ photodiodes, $N$ Faraday rotators, $N$ couplers, and $N$ polarization controllers.

\begin{table*}[t]
\caption{Control-Path Hardware for Phase Control of $N$ Beams (Combiner, Coherent Receiver, and Front-End EDFAs Excluded) and Simulated Performance}
\label{tab:hw}
\centering\footnotesize
\setlength{\tabcolsep}{4pt}
\renewcommand{\arraystretch}{1.1}
\begin{tabular}{lccccc}
\hline
 & This work, analog & This work, digital & Centralized SPGD & Distributed SPGD & Dither locking \\
\hline
Photodiodes & $2N$ & $2N$ & 1 & $N-1$ & $N-1$ \\
\quad at $N=4/16/64/256$ & 8/32/128/512 & 8/32/128/512 & 1/1/1/1 & 3/15/63/255 & 3/15/63/255 \\
Optical amplifiers & \multicolumn{5}{c}{none in the control path ($N$ front-end EDFAs, common to all schemes)} \\
Readout optics & \multicolumn{2}{c}{$N$ 3-dB couplers, $1\times N$ splitter} & --- & --- & --- \\
Signal-path taps (diverted power, 5\% taps) &$N+1$ (0.45~dB) & $N+1$ (0.45~dB) & 1 (0.22~dB) & 0 (dark ports) & 0 (dark ports) \\
ADC / DAC channels & 0 / 0 ($N$ integrators) & $N$ / $N$ & 1 / $N$ & $N-1$ / $N-1$ & $N-1$ / $N-1$ \\
Perturbation of the beams & none & none & $\pm\delta$, all beams & $\pm\delta$, every stage & sinusoidal, every stage \\
Dither frequencies & 0 & 0 & 0 & 0 & $\log_2N$ \\
Measurements per sample & $N$ gradients & $N$ gradients & 1 scalar & $N-1$ scalars & $N-1$ scalars \\
Time to $\eta=0.9$, $N=4/64$ (simulated) & \multicolumn{2}{c}{0.52/0.54~$\mu$s} & 6.2/32~$\mu$s & 4.6/7.9~$\mu$s & 8.5/18~$\mu$s \\
Residual loss, $N=4/64$ (simulated) & \multicolumn{2}{c}{$3.7\times10^{-4}$/$8.6\times10^{-4}$~dB} & 0.018/0.089~dB & 0.0074/0.021~dB & 0.013/0.038~dB \\
\hline
\multicolumn{6}{l}{Simulated values: $\Omega=10^{3}$~rad/s, OSNR 6~dB per beam, $f_s=100$~MS/s, scaling study (Section~\ref{subsec:simn});}\\
\multicolumn{6}{l}{\quad the digital loop was simulated, the analog loop is its small-step limit.}\\
\multicolumn{6}{l}{Polarization loops (this work only) add $2N$ photodiodes, $N$ Faraday rotators, $N$ couplers, and $N$ polarization controllers.}\\
\hline
\end{tabular}
\end{table*}

\begin{table*}[!tp]
\caption{Settings Selected by the Tuning (Two Training Realizations; Ideal Devices at 100~MS/s; Nominal OSNR 6~dB per Beam unless Stated; Metric: Mean Efficiency over the First 1~ms after Loop Closure)}
\label{tab:tuned}
\centering\footnotesize
\setlength{\tabcolsep}{3.5pt}
\begin{tabular}{rr|c|ccc|ccc|cccc}
\hline
 & & This work & \multicolumn{3}{c|}{Centralized SPGD} & \multicolumn{3}{c|}{Distributed SPGD} & \multicolumn{4}{c}{Dither locking} \\
$N$ & $\Omega$ (rad/s) & $\mu K_{\rm pd}$ ($f_c$, MHz) & $f_{\rm it}$ (MHz) & $\delta$ (rad) & $\kappa$ & $f_{\rm it}$ (MHz) & $\delta$ (rad) & $\kappa$ & $\phi_{\rm step}$ ($^{\circ}$) & $\phi_{\rm mag}$ ($^{\circ}$) & $f_{d,\max}$ (MHz) & $m$ \\
\hline
\multicolumn{13}{l}{\emph{Rate study}}\\
4 & 10 & 0.272 (4.32) & 5 & 0.05 & 0.2 & 5 & 0.03 & 0.3 & 1.64 & 3.84 & 12.5 & 1 \\
4 & 30 & 0.272 (4.32) & 5 & 0.05 & 0.2 & 5 & 0.03 & 0.3 & 1.64 & 3.84 & 12.5 & 1 \\
4 & 100 & 0.272 (4.32) & 5 & 0.05 & 0.2 & 5 & 0.03 & 0.3 & 1.64 & 3.84 & 12.5 & 1 \\
4 & 300 & 0.272 (4.32) & 5 & 0.05 & 0.2 & 5 & 0.03 & 0.3 & 1.64 & 3.84 & 12.5 & 1 \\
4 & $10^{3}$ & 0.272 (4.32) & 5 & 0.05 & 0.2 & 5 & 0.03 & 0.3 & 1.64 & 3.84 & 12.5 & 1 \\
4 & $3\times10^{3}$ & 0.384 (6.11) & 10 & 0.05 & 0.1 & 5 & 0.03 & 0.3 & 1.64 & 3.84 & 12.5 & 1 \\
4 & $10^{4}$ & 0.272 (4.32) & 5 & 0.05 & 0.2 & 5 & 0.03 & 0.3 & 1.64 & 3.84 & 12.5 & 1 \\
4 & $3\times10^{4}$ & 0.272 (4.32) & 5 & 0.05 & 0.3 & 10 & 0.05 & 0.3 & 2.46 & 3.84 & 12.5 & 1 \\
4 & $10^{5}$ & 0.192 (3.06) & 10 & 0.07 & 0.3 & 10 & 0.05 & 0.5 & 3.28 & 5.76 & 12.5 & 1 \\
\hline
\multicolumn{13}{l}{\emph{Scaling study}}\\
4 & $10^{3}$ & 0.272 (4.32) & 5 & 0.05 & 0.2 & 5 & 0.03 & 0.3 & 1.64 & 3.84 & 12.5 & 1 \\
4 & $10^{5}$ & 0.192 (3.06) & 10 & 0.07 & 0.3 & 10 & 0.05 & 0.5 & 3.28 & 5.76 & 12.5 & 1 \\
8 & $10^{3}$ & 0.384 (6.11) & 10 & 0.07 & 0.3 & 10 & 0.05 & 0.5 & 2.46 & 3.84 & 12.5 & 1 \\
8 & $10^{5}$ & 0.192 (3.06) & 25 & 0.1 & 0.2 & 10 & 0.05 & 0.7 & 3.28 & 5.76 & 12.5 & 1 \\
16 & $10^{3}$ & 0.272 (4.32) & 10 & 0.07 & 0.3 & 10 & 0.03 & 0.2 & 2.46 & 3.84 & 12.5 & 1 \\
16 & $10^{5}$ & 0.543 (8.64) & 25 & 0.1 & 0.3 & 10 & 0.05 & 0.7 & 3.28 & 5.76 & 12.5 & 1 \\
64 & $10^{3}$ & 0.543 (8.64) & 25 & 0.1 & 0.3 & 10 & 0.03 & 0.2 & 2.46 & 2.88 & 12.5 & 1 \\
64 & $10^{5}$ & 0.384 (6.11) & 25 & 0.15 & 0.7 & 25 & 0.05 & 0.3 & 4.92 & 5.76 & 12.5 & 1 \\
\hline
\multicolumn{13}{l}{\emph{OSNR study at constant power} (EDFA output 0~dBm, $\Omega=10^{3}$~rad/s; second column: OSNR per beam in dB, every second point of the 2-dB grid)}\\
4 & 10 & 0.272 (4.32) & 5 & 0.05 & 0.3 & 10 & 0.03 & 0.3 & 2.46 & 2.88 & 12.5 & 1 \\
4 & 6 & 0.272 (4.32) & 5 & 0.05 & 0.2 & 5 & 0.03 & 0.3 & 1.64 & 3.84 & 12.5 & 1 \\
4 & 2 & 0.272 (4.32) & 5 & 0.07 & 0.2 & 5 & 0.05 & 0.2 & 1.64 & 5.76 & 12.5 & 1 \\
4 & $-2$ & 0.0679 (1.08) & 5 & 0.07 & 0.1 & 5 & 0.07 & 0.1 & 0.82 & 5.76 & 12.5 & 3 \\
4 & $-6$ & 0.048 (0.764) & 5 & 0.1 & 0.1 & 5 & 0.1 & 0.05 & 0.82 & 7.67 & 12.5 & 3 \\
4 & $-10$ & 0.017 (0.27) & 5 & 0.1 & 0.05 & 5 & 0.1 & 0.03 & 0.82 & 15.3 & 12.5 & 3 \\
\hline
\multicolumn{13}{l}{\emph{OSNR study at $\Omega=10^{5}$~rad/s} (EDFA output 0~dBm; second column: OSNR per beam in dB)}\\
4 & 10 & 0.272 (4.32) & 25 & 0.07 & 0.2 & 10 & 0.03 & 0.7 & 3.28 & 3.84 & 12.5 & 1 \\
4 & 8 & 0.192 (3.06) & 25 & 0.07 & 0.2 & 10 & 0.05 & 0.7 & 3.28 & 3.84 & 12.5 & 1 \\
4 & 6 & 0.192 (3.06) & 10 & 0.07 & 0.3 & 10 & 0.05 & 0.5 & 3.28 & 5.76 & 12.5 & 1 \\
4 & 4 & 0.192 (3.06) & 10 & 0.07 & 0.3 & 10 & 0.07 & 0.5 & 3.28 & 5.76 & 12.5 & 1 \\
4 & 2 & 0.192 (3.06) & 10 & 0.1 & 0.3 & 10 & 0.07 & 0.5 & 4.92 & 7.67 & 12.5 & 1 \\
4 & 0 & 0.136 (2.16) & 10 & 0.1 & 0.3 & 5 & 0.1 & 0.7 & 3.28 & 7.67 & 12.5 & 3 \\
\hline
\multicolumn{13}{l}{\emph{EDFA output study} (OSNR 6~dB; second column: EDFA output power per beam in dBm)}\\
4 & $-10$ & 0.272 (4.32) & 5 & 0.05 & 0.2 & 5 & 0.03 & 0.3 & 1.64 & 3.84 & 12.5 & 1 \\
4 & 0 & 0.272 (4.32) & 5 & 0.05 & 0.2 & 5 & 0.03 & 0.3 & 1.64 & 3.84 & 12.5 & 1 \\
4 & 10 & 0.272 (4.32) & 5 & 0.05 & 0.2 & 5 & 0.03 & 0.3 & 1.64 & 3.84 & 12.5 & 1 \\
4 & 20 & 0.272 (4.32) & 5 & 0.05 & 0.2 & 5 & 0.03 & 0.3 & 1.64 & 3.84 & 12.5 & 1 \\
\hline
\multicolumn{13}{l}{\emph{High-rate study} (OSNR per beam 7, 1, and $-5$~dB for $N=4$, 16, and 64, EDFA output 0~dBm; every second point of the half-decade grid)}\\
4 & $10^{4}$ & 0.272 (4.32) & 5 & 0.05 & 0.2 & 5 & 0.03 & 0.3 & 1.64 & 2.88 & 12.5 & 1 \\
4 & $10^{5}$ & 0.192 (3.06) & 10 & 0.07 & 0.3 & 10 & 0.05 & 0.7 & 3.28 & 5.76 & 12.5 & 1 \\
4 & $10^{6}$ & 0.768 (12.2) & 25 & 0.15 & 0.7 & 25 & 0.1 & 1.5 & 24.6 & 15.3 & 12.5 & 1 \\
4 & $10^{7}$ & 1.54 (24.4) & 25 & 0.3 & 1.5 & 25 & 0.5 & 3 & 32.8 & 30.7 & 12.5 & 1 \\
16 & $10^{4}$ & 0.272 (4.32) & 10 & 0.1 & 0.3 & 2.5 & 0.05 & 0.5 & 2.46 & 5.76 & 12.5 & 1 \\
16 & $10^{5}$ & 0.543 (8.64) & 25 & 0.15 & 0.3 & 10 & 0.07 & 0.5 & 4.92 & 7.67 & 12.5 & 1 \\
16 & $10^{6}$ & 0.543 (8.64) & 25 & 0.2 & 1 & 25 & 0.15 & 0.7 & 24.6 & 23 & 12.5 & 1 \\
16 & $10^{7}$ & 1.54 (24.4) & 25 & 0.5 & 5 & 25 & 0.5 & 3 & 32.8 & 23 & 12.5 & 1 \\
64 & $10^{4}$ & 0.0679 (1.08) & 25 & 0.15 & 0.2 & 5 & 0.07 & 0.1 & 1.64 & 7.67 & 12.5 & 3 \\
64 & $10^{5}$ & 0.096 (1.53) & 25 & 0.3 & 0.7 & 10 & 0.1 & 0.2 & 6.56 & 15.3 & 12.5 & 1 \\
64 & $10^{6}$ & 0.384 (6.11) & 25 & 0.3 & 5 & 25 & 0.3 & 0.7 & 24.6 & 30.7 & 12.5 & 1 \\
64 & $10^{7}$ & 1.09 (17.3) & 25 & 0.3 & 10 & 25 & 0.5 & 2 & 32.8 & 23 & 12.5 & 1 \\
\hline
\multicolumn{13}{l}{Fading study (Fig.~\ref{fig:cmppower}): every scheme keeps the settings of the rate study at $10^{3}$~rad/s; the scheduled loop uses the effective step of the}\\
\multicolumn{13}{l}{\quad sampled design rule, $\mu K_{\rm pd}=0.0082$, 0.0047, 0.0023, and $7.6\times10^{-4}$ at an OSNR of 6, 0, $-6$, and $-14$~dB.}\\
\multicolumn{13}{l}{$m$: number of demodulated dither periods summed before each sign decision.}\\
\hline
\end{tabular}
\end{table*}

\subsection{Summary of the Comparison}
Under equal conditions---identical disturbances, the same 100-MS/s clock, the same front-end EDFA, ideal devices, the OSNR at which ideal combining reaches the FEC threshold, and every scheme tuned for every condition---the proposed loop had the lowest residual loss of the four controllers at every rms phase rate from 10 to $10^{5}$~rad/s, at every array size from 4 to 64 beams, at every OSNR from $-10$ to 10~dB, and in fades, lower by a factor of 13 to 260 than the retuned baselines in the rate and scaling studies, and it reached $\eta=0.9$ within 0.6~$\mu$s at every rate and array size, 2.8 to 59 times faster than the baselines wherever they held the lock, with one step size that the design rule predicts. Up to $10^{5}$~rad/s the clock does not limit the proposed loop, whose best steady-state step stays at $\mu K_{\rm pd}\le0.19$, but it still bounds the baselines, which need several samples per update---four for SPGD and eight for dither locking---and select the highest rate the clock allows where speed matters. At this clock rate every scheme holds the lock up to $10^{5}$~rad/s for four beams, whereas centralized SPGD loses it at 64 beams and $10^{5}$~rad/s. Beyond $10^{5}$~rad/s the clock bounds every scheme, yet 1~dB above the FEC threshold the proposed loop keeps the BER of ideal combining up to $10^{7}$~rad/s for four, sixteen, and sixty-four beams, whereas every baseline crosses the threshold between $10^{5}$ and $10^{7}$~rad/s, the earlier the more beams, and makes the carrier recovery slip. The other differences lie in the acquisition, which slows with $N$ for centralized SPGD and dither locking (32 and 18~$\mu$s at 64 beams and $10^{3}$~rad/s) but not for the proposed loop; in the noise tolerance, since between $-10$ and 10~dB the proposed loop loses 8 to 100 times less and, in an extended sweep, the baselines fail at an OSNR about 10~dB higher; in fades without retuning, where the losses of the baselines grow to 0.2--0.9~dB at an OSNR of $-14$~dB against $4\times10^{-3}$~dB; and in the perturbation, which every baseline applies to the beams. For a 28-GBd DP-QPSK signal behind the front-end EDFAs, the residual loss of each retuned scheme becomes an equal OSNR penalty at the FEC threshold, at most 0.018~dB at $10^{3}$~rad/s and 0.054~dB at $10^{5}$~rad/s, whereas the readout taps, which divert 0.45~dB (5\% taps) or 0.09~dB (1\% taps) of the combined power, leave the OSNR unchanged; brief misalignments of dither locking cost it more than its residual loss, 0.14~dB at $10^{5}$~rad/s and an OSNR of 10~dB, and 1.0~dB at the threshold with the settings held fixed. Raising the EDFA output beyond 0~dBm per beam changes none of these results. Distributed SPGD came closest throughout, but it perturbs the beams. The proposed scheme pays with $2N$ photodiodes and with readout taps that cost combined power but no SNR.

\section{Assumptions and Limitations}
\label{sec:limits}
\begin{enumerate}
\item \emph{Ideal-port algebra.} The readout identities~\eqref{eq:balance}, \eqref{eq:gammaselect}, and~\eqref{eq:readoutgrad} hold for ideal ports. Splitting-ratio errors, rotator errors, responsivity mismatch, and polarity errors are static: they are absorbed into $K_{\varphi}$, $K_{\theta}$, the polarities, and the offset trims at calibration, and temperature drift requires periodic recalibration against known control states. The identities state an achievable relative accuracy, not the behavior of an uncalibrated device.
\item \emph{Timing pre-alignment.} Coarse delay alignment between branches is assumed; the loops do not recover timing. A residual inter-branch rms delay $\sigma_\tau$ costs approximately $\bigl[8.686\,c(\beta)+4.343\,S(\beta)/N\bigr](\sigma_\tau/T)^{2}$~dB in combining efficiency and intersymbol interference jointly for a raised-cosine pulse of rolloff $\beta$, with $c(0.2)=1.68$ and $S(0.2)=2.38$, so $\sigma_\tau\lesssim0.08\,T$ keeps the penalty below $0.1$~dB for rolloffs $0$--$1$ and $N=4$--$16$. The readout is more forgiving than the combining: a branch--reference delay $\delta$ only scales the measured correlation by the pulse autocorrelation $\rho(\delta)=1-(\delta^{2}/2)\langle p'^{2}\rangle/\langle p^{2}\rangle$, which exceeds $0.99$ for $\delta<0.05\,T$. An adjustable delay in each branch, outside the loop, is the natural remedy.
\item \emph{Readout path matching.} The null of the measured correlation coincides with output alignment only if the path from the branch tap to its coupler equals the path from the reference tap to the same coupler plus the branch's path through the combiner tree, up to a constant common to all units. A differential violation $\Delta$ becomes a static inter-branch phase error that costs $\approx1.1\,\Delta^{2}$~dB for two equal branches, whereas a common violation is absorbed in the unobservable common phase. The condition concerns the carrier phase and is trimmed by the static offset $\gamma$ of each unit; an electronic trim at the branch actuator cannot serve, because it shifts the measurement null and the output phase equally. Sub-$0.1$~dB operation asks for differential matching of order $0.1$~rad, natural on an integrated combiner and requiring an active trim in fiber.
\item \emph{Fading-dependent loop gain.} With fixed gains the loop bandwidth~\eqref{eq:fc} is proportional to the instantaneous branch and reference field product. Deep fades slow the loops, which is stabilizing, while scintillation transients must fit within the bandwidth set at full power; gain headroom, bounded updates, and gating are the safeguards. A digital loop can instead schedule its step with the monitored power, which Section~\ref{subsec:simpower} simulates only for static fades common to all beams; behind the 47-dB front end the readout stays limited by ASE beating over the simulated 20-dB fade, whereas with a front-end gain of only 20~dB the thermal noise of the detectors would dominate already at the nominal power, so the front-end gain sets how deep a fade the readout tolerates. The design rule~\eqref{eq:design}--\eqref{eq:designopt} needs the disturbance and noise levels of the actual link, which are model quantities here.
\item \emph{Restricted polarization model.} A single real rotation per branch cannot compensate general complex differential Jones mixing or polarization-dependent loss, and the geometric picture of Section~\ref{subsec:geometry} is exact only for linear polarizations. Extending the compensation to a general unitary Jones matrix requires more actuators and correspondingly more readout correlations; the present formulation does not cover it.
\item \emph{Common ambiguities.} A common phase and a common polarization rotation are unobservable to the relative loops and are handled by the processing of the single coherent receiver. The $N-1$ effective degrees of freedom of an $N$-input passive combiner are spanned by the branch-resolved loops; the common mode is not.
\item \emph{Equal-ratio combining only.} Without an amplitude actuator, branches are combined at their own field amplitudes. This is the maximal-ratio solution when the branch noises are equal, and the penalty of unequal branch noise, second order in the amplitude spread, is accepted rather than compensated: a deeply faded branch is neither switched off nor down-weighted. The simplification removes $N$ amplitude actuators, $N$ further loops, and the branch-quality statistics that amplitude weighting would need.
\item \emph{No experimental validation.} No laboratory or field demonstration accompanies this work. Combiner insertion loss, the power budget of the monitoring tap, actuator bandwidth, and noise management are analyzed architecturally but not measured. The comparison of Section~\ref{sec:compare} uses a deliberately simplified model---co-polarized beams of equal amplitude, an ideal combiner, detectors without bandwidth limits, a synthetic multi-tone phase disturbance, white detector noise derived from a link budget with a common front-end EDFA, digital controllers on a common 100-MS/s clock, no polarization control, and a bit-error ratio derived from the combining efficiency for an ASE-limited receiver with an ideal front end, timing, polarization demultiplexing, and carrier recovery (Fig.~\ref{fig:dsp}), checked against a symbol-level simulation of the complete chain with blind carrier recovery but without laser phase noise or frequency offset---so it ranks control principles under equal conditions and does not predict the performance of particular devices.
\end{enumerate}

\section{Conclusion}
The scheme combines $N$ field replicas optically, ahead of photodetection, and aligns them by exact gradient ascent on the total combined power. The per-branch gradients with respect to the phase and the polarization angle, $-2\Imag\{\mathbf{s}^{H}\mathbf{v}_i\}$ and $2\Real\{\mathbf{s}^{H}\mathbf{J}\mathbf{v}_i\}$, are read out directly as balanced photocurrents: by a 3~dB coupler for the phase, and by the same coupler behind a $90^{\circ}$ Faraday rotator and a $\pi/2$ offset for the polarization. Geometrically, each readout is the part of the branch field that is out of alignment with the combined field---in quadrature for the phase, along $\mathbf{J}\mathbf{v}_i$ for the polarization---so each fixed-gain loop turns its branch toward the combined field and comes to rest when the branch is aligned; near alignment the two loops of a branch are independent first-order loops, and far from it the phase loop has to lock first. The reference is a passive tap of the combined field, every branch carries only a phase and a polarization actuator, all branches update simultaneously, and the loops can be analog integrators or a low-rate processor with no division. In a phase-only simulation under equal conditions, with the same front-end amplifier and the same 100-MS/s clock for all schemes and a per-beam OSNR at the FEC threshold, the digital form of the loop had the lowest residual loss of four controllers for rms phase rates from 10 to $10^{5}$~rad/s, up to 64 beams, at every OSNR from $-10$ to 10~dB, and in fades, acquired within 0.6~$\mu$s at every rate and array size, held the lock where centralized stochastic parallel gradient descent lost it, kept the bit-error ratio of ideal combining 1~dB above the FEC threshold up to $10^{7}$~rad/s for four, sixteen, and sixty-four beams, where every baseline failed, and needed only the step size that the design rule predicts. It pays with $2N$ photodiodes and with readout taps that divert combined power but, behind the front-end amplifiers, leave the OSNR and the bit-error ratio of a DP-QPSK signal unchanged; its fixed gain lets its bandwidth fall with the received power, yet with the settings held fixed it keeps its loss below $5\times10^{-3}$~dB down to an OSNR of $-14$~dB per beam, where the baselines lose 0.2 to 0.9~dB. The assumptions---ideal port algebra with calibration, timing pre-alignment, readout path matching, a fading-dependent fixed gain, a restricted polarization model, and equal-ratio combining---are stated explicitly, and experimental validation remains open.

\section*{Code and Data Availability}
The arXiv source bundle contains only the manuscript, bibliography, class and style files, and figures; public distribution of simulation code and patent-derived material requires the authors' clearance. No public repository or submission identifier is asserted in this draft.

\bibliographystyle{IEEEtran}
\bibliography{references}
\end{document}